\documentclass[aps,floats]{revtex4}
\usepackage{amsmath,amssymb}
\usepackage{graphicx,epsfig}

\begin{document}
\bibliographystyle {plain}

\def\oppropto{\mathop{\propto}} 
\def\opsimeq{\mathop{\simeq}}
\def\opoverderline{\mathop{\overline}}
\def\operarrow{\mathop{\longrightarrow}}
\def\opsim{\mathop{\sim}}

\def\fig#1#2{\includegraphics[height=#1]{#2}}
\def\figx#1#2{\includegraphics[width=#1]{#2}}


\title{ Freezing transition of the random bond RNA model: \\
statistical properties of the pairing weights   } 


\author{ C\'ecile Monthus and Thomas Garel }
 \affiliation{Service de Physique Th\'{e}orique, CEA/DSM/SPhT\\
Unit\'e de recherche associ\'ee au CNRS\\
91191 Gif-sur-Yvette cedex, France}

\begin{abstract}

To characterize the pairing-specificity of RNA secondary structures
as a function of temperature, we analyse the statistics of the pairing weights as follows :
 for each base $(i)$ of the sequence of length N,  
we consider the $(N-1)$ pairing weights $w_i(j)$
with the other bases $(j \neq i)$ of the sequence. 
We numerically compute the probability distributions 
$P_1(w)$ of the maximal weight $w_i^{max}= max_j [w_i(j)]$, 
the probability distribution $\Pi(Y_2)$ of the parameter
$Y_2(i)= \sum_j w_i^2(j)$, as well as the average values of the moments
$Y_k(i)= \sum_j w_i^k(j)$. 
We find that there are two important temperatures $T_c<T_{gap}$.
For $T>T_{gap}$, the distribution $P_1(w)$ vanishes at some value $w_0(T)<1$,
and accordingly the moments $\overline{Y_k(i)}$ decay exponentially 
as $(w_0(T))^k$ in $k$.
For $T<T_{gap}$, the distributions $P_1(w)$ and $\Pi(Y_2)$
present the characteristic Derrida-Flyvbjerg 
singularities at $w=1/n$ and $Y_2=1/n$
for $n=1,2..$. In particular, there exists a temperature-dependent
exponent $\mu(T)$ that governs the singularities 
$P_1(w) \sim (1-w)^{\mu(T)-1}$ and $\Pi(Y_2) \sim (1-Y_2)^{\mu(T)-1}$
as well as the power-law decay of the
 moments $ \overline{Y_k(i)} \sim 1/k^{\mu(T)}$.
 The exponent $\mu(T)$ grows from the value $\mu(T=0)=0$ up to $\mu(T_{gap}) \sim 2$.
The study of spatial properties indicates that the critical temperature $T_c$
where the large-scale roughness exponent changes from the low temperature
value $\zeta \sim 0.67$ to the high temperature value $\zeta \sim 0.5$
corresponds to the exponent $\mu(T_c)=1$. 
For $T<T_c$, there exists frozen pairs of all sizes, 
whereas for $T_c< T <T_{gap}$, there exists frozen pairs, but only up to some
characteristic length diverging as $\xi(T) \sim 1/(T_c-T)^{\nu}$ with
$\nu \simeq 2$.
The similarities and differences with the weight statistics in L\'evy
sums and in Derrida's Random Energy Model are discussed.

\bigskip


\end{abstract}

\maketitle
\section{ Introduction}

Models of RNA secondary structures \cite{Higgs_review,Sch}
have been recently studied by physicists 
\cite{Higgs96,Mor_Hig,Krz_Mez_Mul,Bun_Hwa,
Pag_Par_Ric,Mar_Pag_Ric,Bur_Har,Wie,Hui_Tan},
because of the similarity of the low-temperature disorder-dominated
phase with the spin glass phase.
This phase has been analysed either in
terms of replica theory of spin-glasses \cite{replica},
with non-trivial overlap distribution function $P(q)$ 
\cite{Higgs96,Pag_Par_Ric,Bur_Har}, or in terms of
the droplet theory of spin-glasses \cite{Fis_Hus_SG} :
in this case, 
 a finite droplet exponent $\theta>0$ has been obtained
via the $\epsilon$-coupling method, with finite values
$\theta \sim 0.23$ \cite{Krz_Mez_Mul,Bur_Har} , $\theta=1/3$ \cite{Mar_Pag_Ric}
whereas a vanishing droplet exponent $\theta=0$ and
 logarithmic droplets have been found via the statistics
of pinch free energies\cite{Bun_Hwa,Hui_Tan}.
Many authors have also studied the phase transition
towards the high-temperature molten phase, with different values
for the correlation length exponent $\nu$ and the specific heat
exponent $\alpha=2-\nu$. Numerical results
have given for instance $\nu \sim 3.9$ 
\cite{Pag_Par_Ric} or $\nu \sim 1.1 $ \cite{Bur_Har},
and the field theory of \cite{Wie} predicts $\nu \sim 8/5 \sim 1.6$,
whereas the general theorem of \cite{chayes} on phase transitions
in disordered systems states that the finite-size
correlation exponent $\nu$ has to satisfy the bound $\nu \geq 2/d=2$. 

In this paper, we try to clarify the nature of the low temperature
phase and of the freezing transition by studying 
statistical properties of the pairing weights.

The paper is organized as follows.
The model and usual observables are recalled in Section \ref{model}.
We then present a detailed study of the pairing weights seen by a
given monomer. For clarity, the statistical properties of the weights
alone, independently of the distances involved are described
in Section \ref{poids}, whereas the study of spatial properties
is given in Section \ref{spatial}. We summarize our results in Section
\ref{conclusion}. For comparison, we recall in the Appendix 
the properties of the weights statistics in L\'evy sums and in Derrida's 
Random Energy Model (REM), as well as the corresponding
Derrida-Flyvberg singularities.

\section{ Model and observables }

\label{model}

\subsection{ Partition function}

An RNA secondary structure of a sequence of $N$ bases $(1,2,...,N)$
is a set of base pairs all compatible with each other.
 To be compatible, two pairs $(i,j)$ and $(k,l)$
have to be non-overlapping (for instance $i<j<k<l$) or
nested (for instance $i<k<l<j$) \cite{Higgs_review}
The energy of an allowed configuration $\cal C$ is then
the sum of the energies $ \epsilon_{i,j}$ of all the pairs
$(i,j)$ that are present in the configuration
\begin{equation}
E({\cal C}) = \sum_{(i,j) \in {\cal C} } \epsilon_{i,j}
\end{equation}

This non-crossing property of pairs allows to write
the following recursion for the partition functions $Z_{i,j}$
of intervals $(i,i+1,...j-1,j)$ \cite{Higgs_review}
\begin{equation}
Z_{i,j}= Z_{i,j-1} + \sum_{k=i}^{j-p_{min}}
 Z_{i,k-1} e^{-\beta \epsilon_{k,j}}
Z_{k+1,j-1}
\label{recursion}
\end{equation}
The first terms represent the configurations where $j$ is unpaired,
whereas the second term represents the configurations where $j$ is paired
with the base $k \in \{i,i+1,..,j-p_{min} \}$, and $p_{min}$
represents the minimal distance along the sequence to form a pair.
So the full partition function $Z_{1,N}$ can be computed 
in a CPU time of order $O(N^3)$.
In the literature \cite{Higgs96,Mor_Hig,Krz_Mez_Mul,Bun_Hwa,
Pag_Par_Ric,Mar_Pag_Ric,Bur_Har,Wie,Hui_Tan} , 
various choices for the parameter $p_{min}$
and for the distribution of the energies $\epsilon_{i,j}$
have been made that we will not rediscuss here. 
All the numerical results we will present below corresponds
to the case $p_{min}=1$ ( the convergence 
towards the asymptotic regime $N \to \infty$ is then much more rapid
than for the `biological value' $p_{min}=4$),
and to the bond-disorder case, where the $\epsilon_{i,j}$ are 
independent random variable drawn with the flat distribution
\begin{equation}
\rho( \epsilon ) = 1 \ \ for \ \  -2.5 \leq \epsilon \leq -1.5
\label{disorder}
\end{equation}
The sequence length $N$ and the corresponding number $ns(N)$ of independent
sequences that we have studied are typically as follows
\begin{eqnarray}
N && =50,100,200,400,600,800 \nonumber \\
ns(N) && =18.10^6, 2.10^6, 3.10^5, 25.10^3,6.10^3, 2.10^3
\label{numerics}
\end{eqnarray}

\subsection{ Pair probabilities }

The pairing probability of bases $(i,j)$ in the sequence $(1,N)$
reads
\begin{equation}
P_{i,j}= \frac{ e^{-\beta \epsilon_{i,j}}  Z_{i+1,j-1} Z_{i,j}^{ext} }{Z_{1,N}}
\label{pairij}
\end{equation}
where $Z_{i+1,j-1}$ represents the partition function
of the internal sequence $(i+1,...,j-1)$ computed in Eq. \ref{recursion},
and where $Z_{i,j}^{ext}$ represents the partition function
of the external sequence $(1,2,..,i-1,j+1,...N)$, which can be computed
by extending the recursion of Eq. \ref{recursion}
to the duplicated sequence $(1,2,..,N,1,2,..N)$:
$Z_{i,j}^{ext}$ is then given by $Z_{j+1,N+i-1}$ \cite{Bun_Hwa}.
So the pair probabilities $P_{i,j}$ can also be computed 
in a CPU time of order $O(N^3)$.

\subsection{ Height profile }

An RNA secondary structure $\cal C$ can be represented 
as a non-crossing arch diagram or equivalently
as a `mountain profile' (see Fig. 3 of \cite{Bun_Hwa})
where the height $h_k$ represents the number of pairs $(i,j)$
such that $i<k<j$ : this height starts at $h(k=1)=0$, ends at $h(k=N)=0$,
remains non-negative in between, and the difference
$(h_{k+1}-h_k)$ can only take the three values $(+1,0,-1)$.
Its thermal average reads in terms of the pair probabilities of Eqs.
(\ref{pairij})
\begin{equation}
<h_k> = \sum_{i<k < j} P_{i,j}
\label{hk}
\end{equation}

\subsection{ Overlap }

In disorder-dominated phases, such as spin-glasses 
or directed polymers for instance, the overlap is usually
a convenient order parameter.
Here, the overlap between two-configurations ${\cal C}_1$
and ${\cal C}_2$ can be defined as 
\begin{equation}
q({\cal C}_1,{\cal C}_2)= \frac{2}{N} \sum_{i<j} 1_{(i,j) \in {\cal C}_1} 
  1_{(i,j) \in {\cal C}_2} 
\label{q2def}
\end{equation}
where the normalization factor $N/2$ represents the number of
pairs existing in the ground state.
The thermal average reads
\begin{equation}
q_2(T) = <q({\cal C}_1,{\cal C}_2)> 
= \frac{2}{N} \sum_{i<j} P_{i,j}^2 
\label{q2}
\end{equation}
in terms of the pair probabilities of Eq. (\ref{pairij}).

However, this overlap is not an appropriate order parameter
for RNA secondary structure, because it does never vanish,
 even at $T=\infty$ as we now explain.

\subsection{ Limit of infinite temperature }

In the limit $T=\infty$, disorder disappears, and the partition function
can be exactly computed \cite{Bun_Hwa}
\begin{equation}
Z_{1,N}(T=\infty) \oppropto_{N \to \infty} 
 \frac{  3^N }{N^{3/2}}
\label{tinfty}
\end{equation}
This number of possible configurations 
corresponds to the number of 1D-random walk for the height $h_k$,
with $3$ choices per step for the height 
increment $h_{k+1}-h_k = 0,\pm 1$, and where 
the factor $1/N^{3/2}$ is a first-return probability. 
From this interpretation of the height as a positive random walk,
it is clear that the middle height $h_{N/2}$ scales as $N^{1/2} $
\begin{equation}
<h_{\frac{N}{2}}> (T=\infty)  \sim N^{1/2}
\label{htinfty}
\end{equation}
For $1 \ll l \ll N$, the pair probability of Eq. \ref{pairij} behaves as
\begin{equation}
P_{i,i+l} (T=\infty) \propto \frac{ N^{3/2} }{ l^{3/2} (N-l)^{3/2} }
\label{pairijtinfty}
\end{equation}
However, the overlap (Eq. \ref{q2}) is finite
\begin{equation}
q_2(T=\infty) >0
\end{equation}
because small pairs have a finite weight, in particular
for $l=1$, Eq. \ref{pairij} yields
\begin{equation}
P_{i,i+1}^{(T=\infty)} \sim \frac{ Z_{1,N-2}(T=\infty) }
{ Z_{1,N} (T=\infty) }
\sim \frac{1}{3^2}
\label{pl1tinfty}
\end{equation}

\subsection{ Limit of zero temperature }

At $T=0$, there is a numerical consensus
\cite{Bun_Hwa,Krz_Mez_Mul,Hui_Tan} that the 
disorder-averaged height has a different scaling from the random walk
value of Eq. \ref{htinfty} 
\begin{equation}
\overline{ <h_{\frac{N}{2}}>} (T=0)   \sim N^{\zeta}
\label{ht0}
\end{equation}
where the roughness exponent 
\begin{equation}
\zeta \sim 0.67
\label{zetat0}
\end{equation}
is extremely close to the simple value $2/3$, although we are not aware
of any rigorous or heuristic argument in favor of this fraction. 
The exponent $\eta$ governing the scaling of large pairs $1 \ll l \ll N$
\begin{equation}
\overline{ P_{i,i+l} (T=0) }  \sim  \frac{1}{N^{\eta}}
 \Phi \left( \frac{l}{N} \right) 
\label{pijt0}
\end{equation}
is actually directly related to the roughness exponent
via the relation \cite{Bun_Hwa,Krz_Mez_Mul,Hui_Tan,Wie}
\begin{equation}
\eta=2 -\zeta
\label{etazeta}
\end{equation}
as can be seen from the definition of the height (Eq. \ref{hk}).
This relation valid at any temperature corresponds to
$\eta(T=\infty)=3/2$ in agreement with Eq. \ref{pairijtinfty},
and to 
\begin{equation}
\eta(T=0) \sim 1.33
\end{equation}
as directly measured in \cite{Krz_Mez_Mul,Hui_Tan}.

\subsection{ Characterization of the transition in previous works }

Previous work has shown that there exists a
finite temperature transition between a high temperature or molten phase,
where entropy dominates, and a low-temperature phase
where disorder dominates. But very different observables have been used
numerically to characterize the transition, such as the overlap
probability distribution $P(q)$ \cite{Higgs96,Pag_Par_Ric,Bur_Har},
the $\epsilon$-coupling method \cite{Krz_Mez_Mul}, and
the so-called `pinch-free energy' \cite{Bun_Hwa,Hui_Tan}.
In the field theory of \cite{Wie}, the critical exponents exactly 
at criticality were found to be the same as the ones in the low
temperature phase, both for the height 
\begin{eqnarray}
\zeta(T > T_c) && = \frac{1}{2} \nonumber \\
\zeta(T \leq T_c) && = \zeta(T=0) 
\label{zetaWie}
\end{eqnarray}
and for the overlap $\overline{ P_{i,i+l}^2 }$ of large pairs $l \gg 1$
\begin{eqnarray}
\overline{ P_{i,i+l}^2 }
&&  \sim \left( \overline{ P_{i,i+l} } \right)^2 \sim \frac{1}{l^3}
\ \ {\rm  for} \ \ T >T_c  \nonumber \\
\overline{ P_{i,i+l}^2 } && \sim \overline{ P_{i,i+l} } 
\sim \frac{1}{l^{\eta(T=0)}} \ \ {\rm for} \ \ T \leq T_c
\label{p2Wie}
\end{eqnarray}

In the following, we propose to study the freezing transition via the 
statistical properties of the pair weights seen by a given monomer.
In \cite{Mor_Hig}, the integrated probability distribution of the maximal weight $p_{max}$
seen by a given monomer has been measured to characterize the barrier
statistics between degenerate ground states for discrete disorder,
but the phase transition region was not studied in details from this point of view.
In another context concerning disordered polymers
\cite{De_Gr_Hi}, the distribution of the maximal weight $p_{max}$ was used to analyse
a phase transition, the important region being there the neighborhood of $p_{max} \to 0$,
whereas in the present study, the important region is the region $p_{max} \to 1$.

\section{ Statistical properties of the pair weights }
\label{poids}

\subsection{Pair landscape seen by a given monomer}

\begin{figure}[htbp]
\includegraphics[height=6cm]{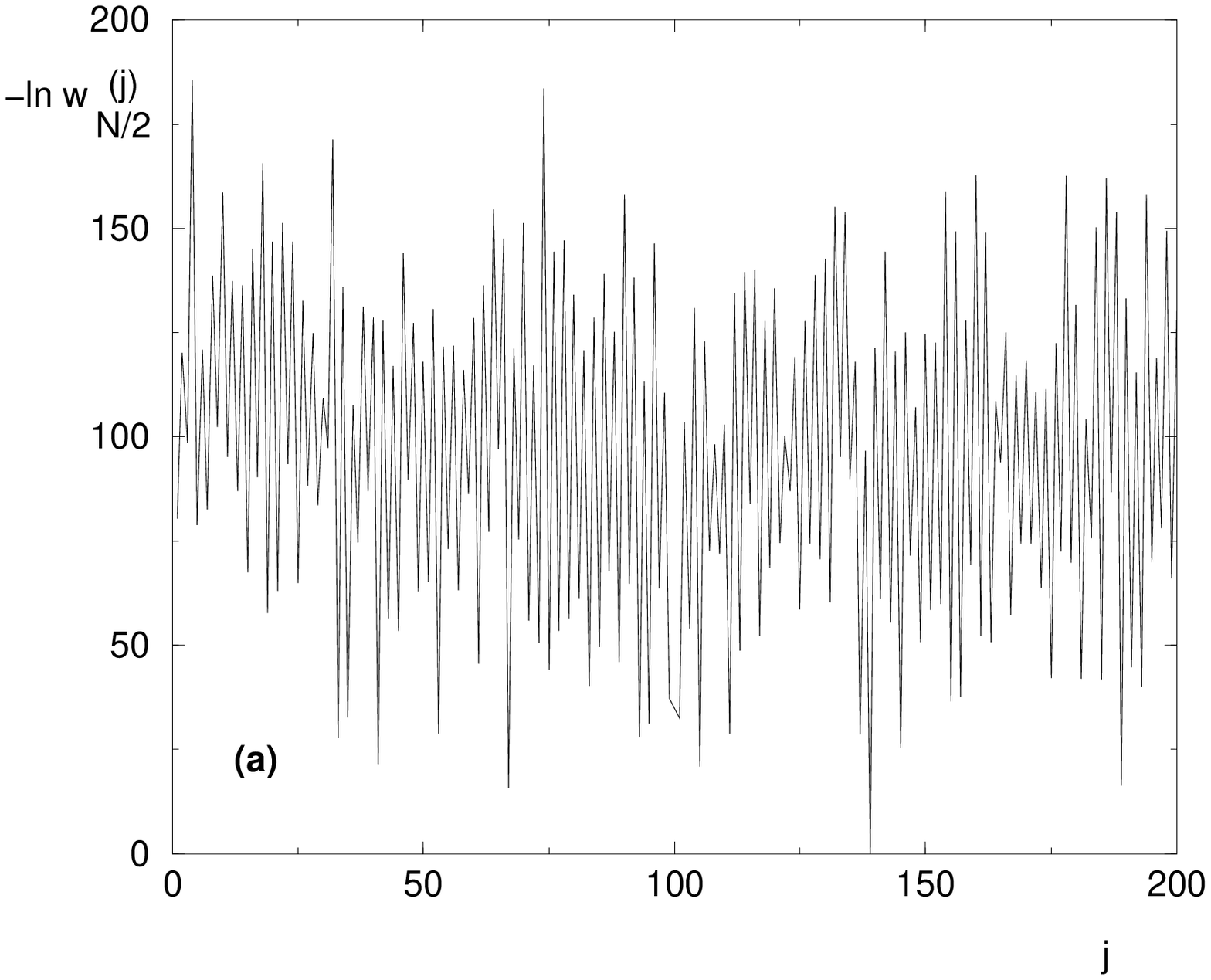}
\hspace{1cm}
\includegraphics[height=6cm]{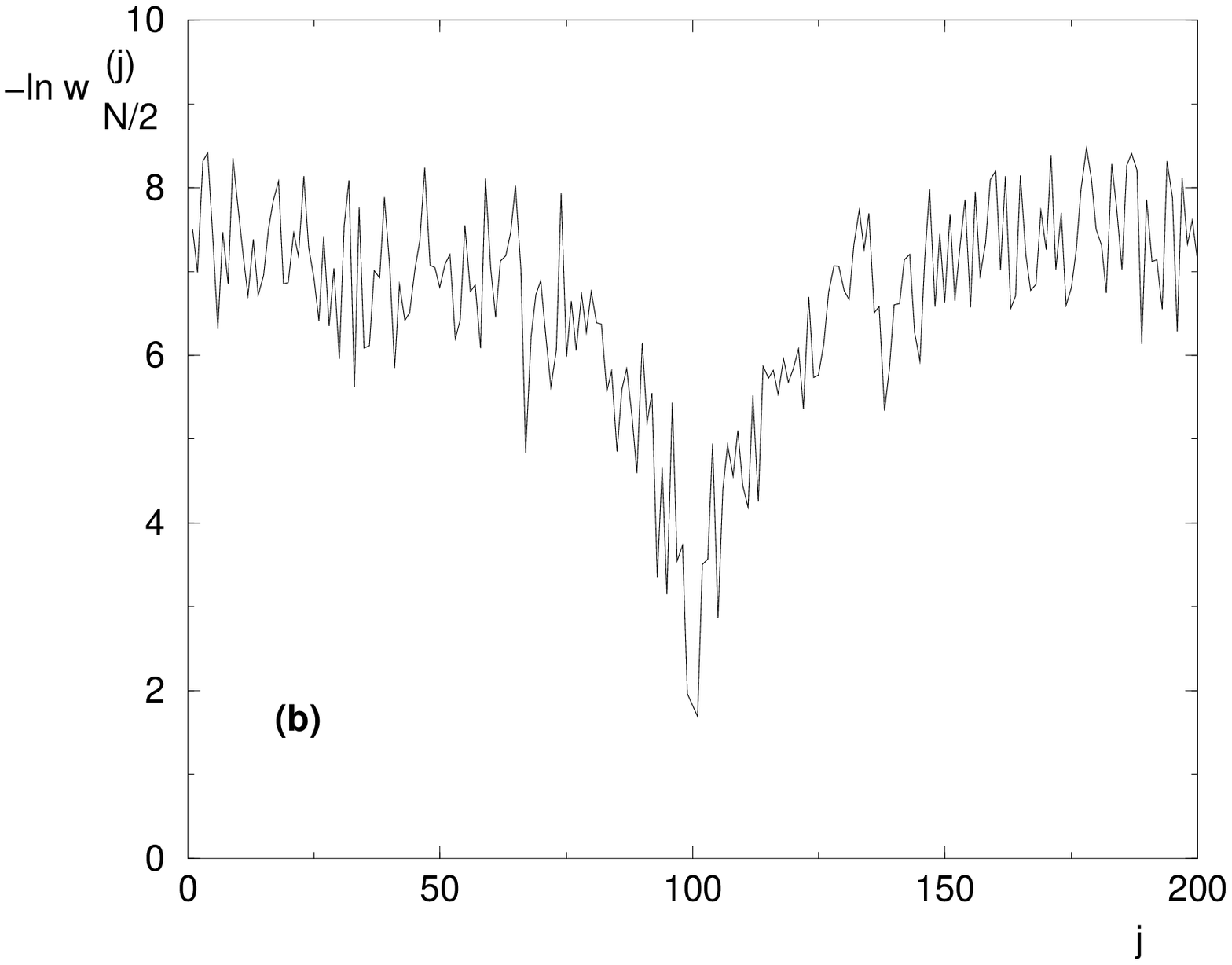}
\caption{(Color online) Pairing weight landscape $(- \ln w_i(j))$ 
seen by the middle monomer
$i=\frac{N}{2}$ for $N=200$ at low and high temperatures
(a) at low temperature ( $T=0.02$ ), only a few weights dominate ( note
the scale of $\ln w_i(j)$ ), at some random positions
  (b) at high temperature ($T=0.5$), the disorder
is a small perturbation with respect to the entropy of the pure case
that favors the neighbors. }
\label{figlandscape}
\end{figure}

For each base $(i)$ of the sequence of length N, 
we consider the $(N-1)$ pairing weights
with the other bases $(j \neq i)$ of the sequence
 (Eq. \ref{pairij})
\begin{equation}
w_i(j) \equiv 
P_{i,j}= \frac{ e^{-\beta \epsilon_{i,j}}  Z_{i+1,j-1} Z_{i,j}^{ext} }{Z_{1,N}}
\end{equation}
Making the convention that $w_i(i)$ denotes the weight of the configurations
where $(i)$ is unpaired 
\begin{equation}
w_i(i) \equiv \frac{ Z_{i,i}^{ext} }{Z_{1,N}}
\end{equation}
these weights are normalized to 
\begin{equation}
\sum_{j \neq i}  w_i(j) =1 -w_i(i)
\end{equation}
The pairing weight landscape
seen by a given monomer is shown for two temperatures on
Fig. \ref{figlandscape}: in the
low-temperature frozen phase, only a few weights dominate 
in continuity with 
  the limit of zero temperature where only one weight is non-zero,
whereas in the high temperature phase, many weights contribute, and
disorder represents a small random correction around the entropic term
of the $T=\infty$ limit (Eq. \ref{pairijtinfty}). 

In the rest of this section, we describe the statistical properties of the weights
alone, independently of the distances involved. The study of spatial properties is
postponed to Section \ref{spatial} for clarity.

\subsection{Characterization of the weights statistics}

In analogy with the weight statistics in L\'evy sums and in
the Random Energy Model \cite{Der,Der_Fly} (we refer the reader to the Appendix
for a summary of the most important results for our present study),
we have numerically computed the probability distributions 
$P_1(w)$ of the maximal weight
\begin{equation}
w_i^{max}= max_{j \neq i} \{ w_i(j) \}
\label{wmax}
\end{equation}
as well as $P_2(w)$ of the second maximal weight.

Another useful way to characterize the statistical
properties of the weights \cite{Der,Der_Fly} (see Appendix)
is to consider the moments of arbitrary order $k$
\begin{equation}
Y_k(i)  = \sum_{j \neq i} w_i^k(j)
\label{ykrna}
\end{equation}
that represents the probability that the monomer $(i)$
is paired to the same monomer in $k$ different thermal configurations
of the same disordered sample.
We have measured the probability $\Pi(Y_2)$ of the parameter
\begin{equation}
Y_2(i)= \sum_{j \neq i} w_i^2(j)
\label{y2rna}
\end{equation}
 as well as the moments $\overline{Y_k(i)}$ for $2 \leq  k \leq 100$.
Finally, we have also computed 
the density of weights
\begin{equation}
f(w) = \overline{ \sum_{j \neq i} \delta(w-w_i(j)) }
\label{densityfrna}
\end{equation}
giving rise to the moments
\begin{equation}
\overline{Y_k(i)} = \int_0^1 dw w^k f(w)
\label{ykf}
\end{equation}
The normalization condition for the density $f(w)$ is
\begin{equation}
\overline{Y_1(i)} = \int_0^1 dw w f(w) =1
\label{normay1}
\end{equation}
The properties of all these quantities in the case of L\'evy sums of
independent variables are recalled in the Appendix. In the following,
we describe  their properties for the RNA case and discuss the
similarities and differences with L\'evy sums. 
The numerical results for the histograms $P_1(w)$,
$P_2(w)$, $f(w)$, $\Pi(Y_2)$ have been obtained by collecting the
weights seen by each monomer $i=1,2..N$ in the $n_s(N)$ disordered
sequences generated (see Eq. \ref{numerics} for the numerical values
used for $N$ and $n_s(N)$).

\subsection{Probability distribution $P_1(w)$
 of the largest weight seen by a given monomer}

\begin{figure}[htbp]
\includegraphics[height=6cm]{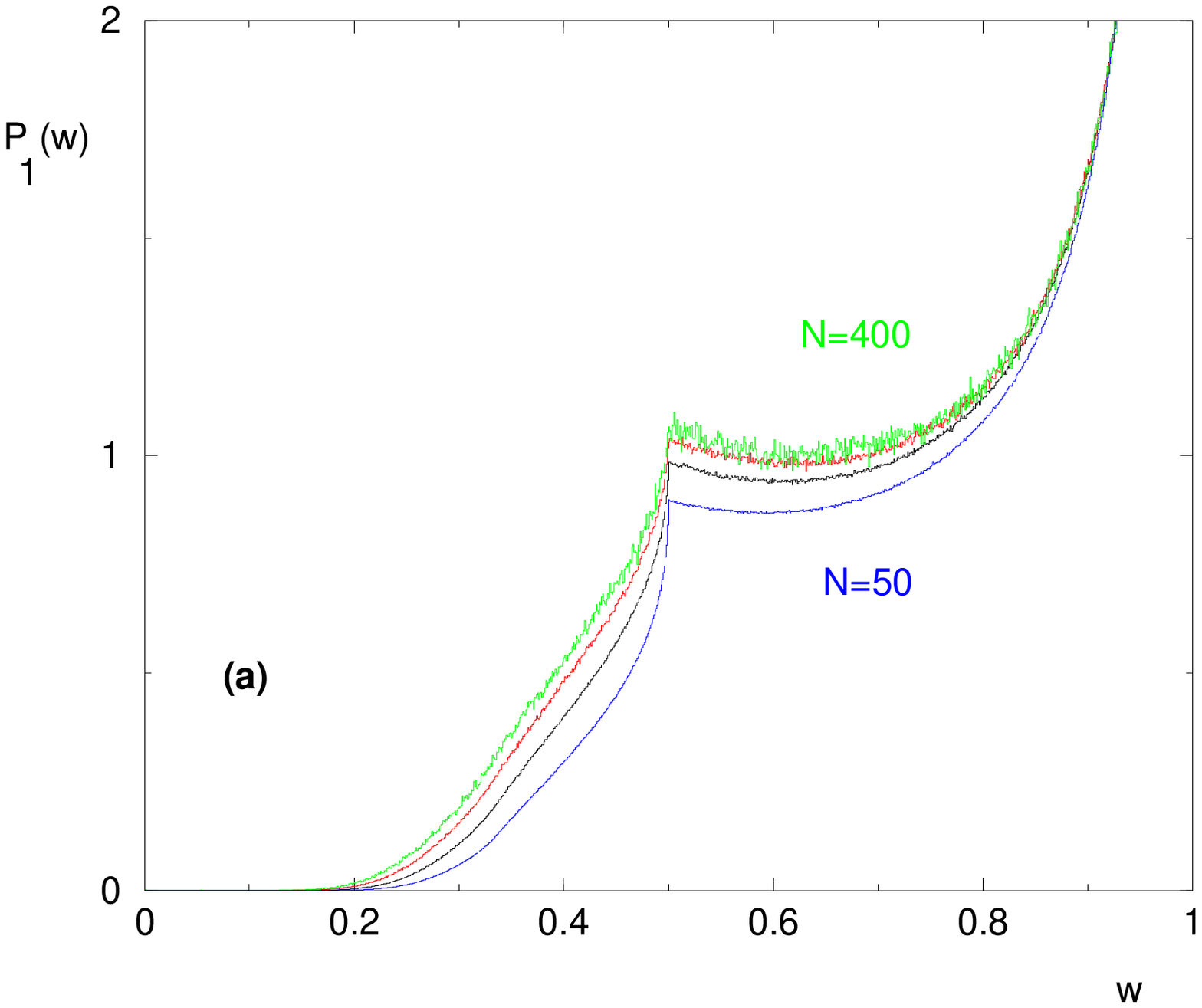}
\hspace{1cm}
\includegraphics[height=6cm]{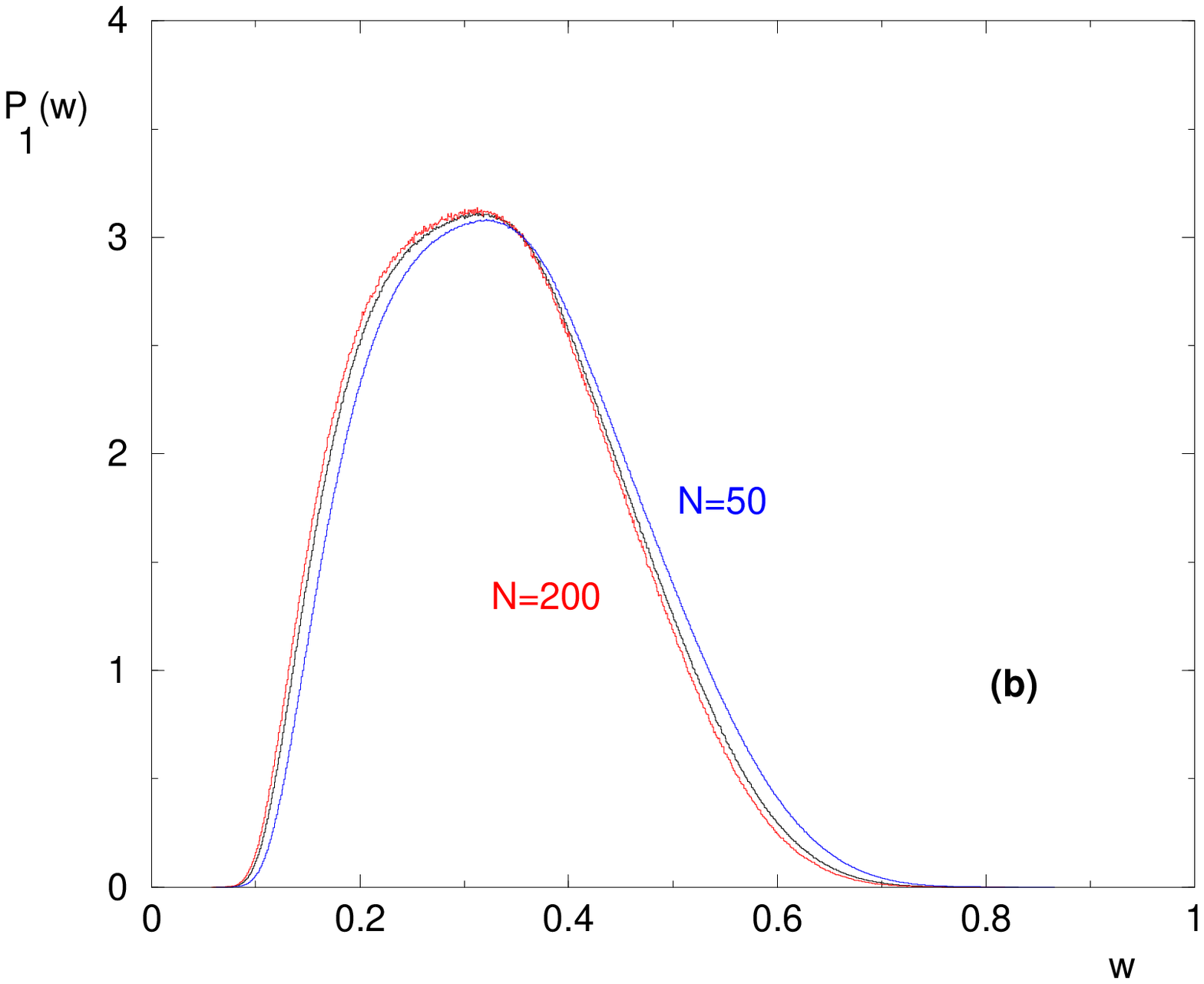}
\caption{(Color online) Probability distribution $P_1(w)$
 of the largest weight seen by a given monomer (see Eq \ref{wmax})
(a) at $T=0.05$ (low-temperature phase) for $N=50,100,200,400$ :
the characteristic Derrida-Flyvbjerg 
singularities at $w=1$ and $w=1/2$ are clearly visible.
 (b) at $T=0.4$ (high-temperature phase) for $N=50,100,200$ :
the distribution $P_1(w)$ does not reach $w=1$ anymore, but vanish
at some maximal value $w_0(T)<1$.  }
\label{figp1w}
\end{figure}

\begin{figure}[htbp]
\includegraphics[height=6cm]{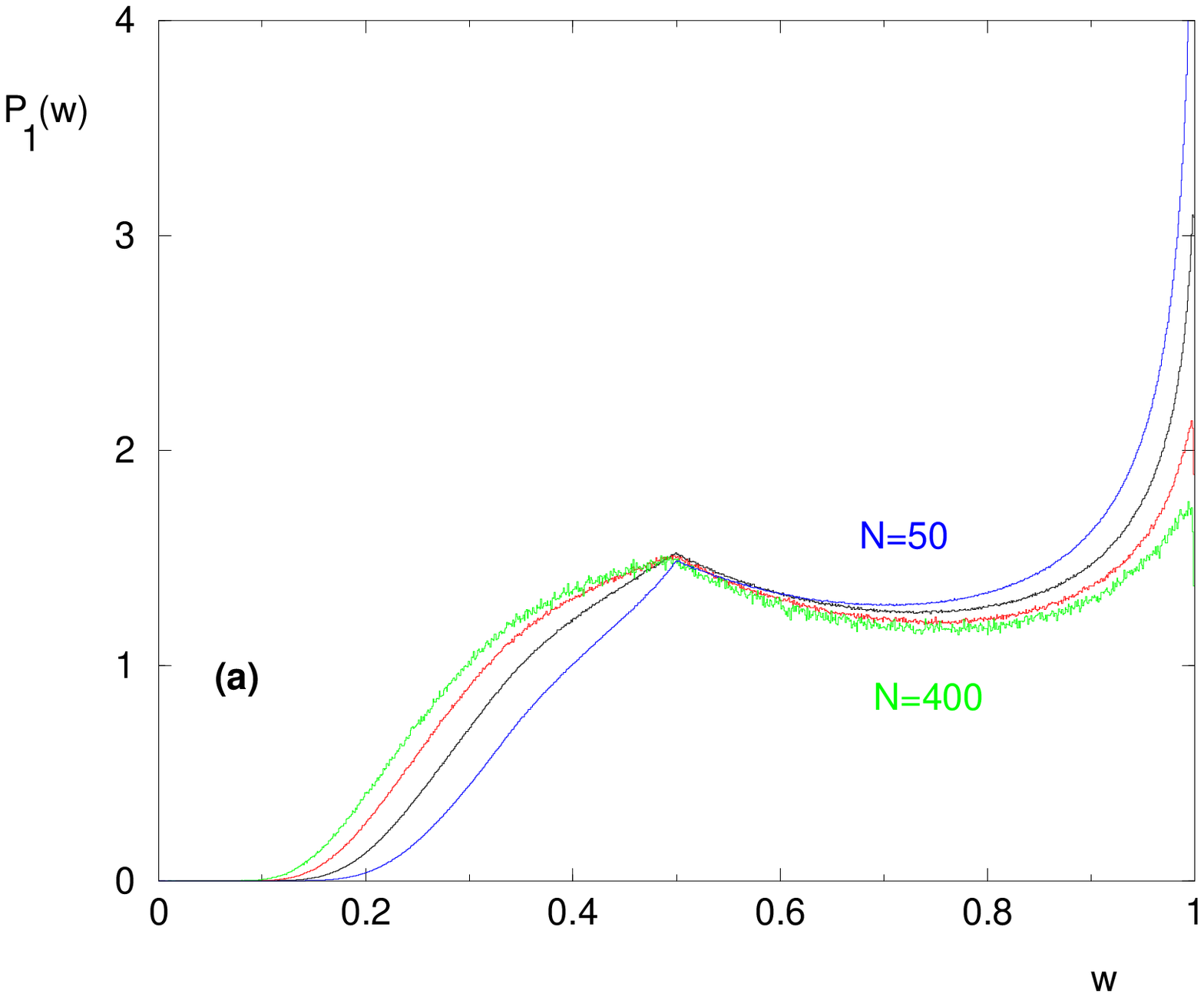}
\hspace{1cm}
\includegraphics[height=6cm]{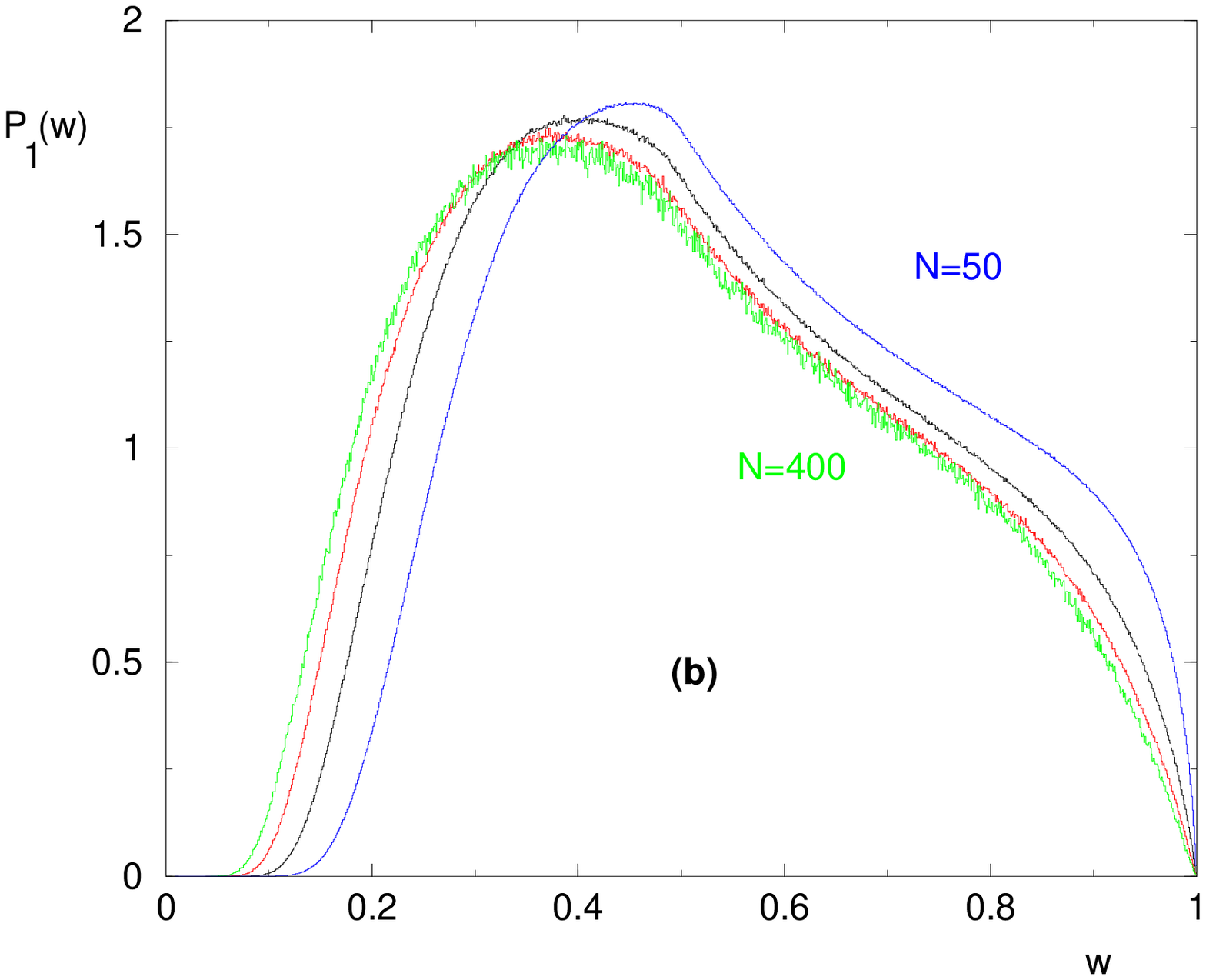}
\caption{(Color online) Probability distribution $P_1(w)$
 of the largest weight seen by a given monomer (see Eq \ref{wmax})
(a) at $T_1 \sim 0.095$ where $\mu(T_1)=1$ for $N=50,100,200,400$ :
$P_1(w)$ does not diverge anymore as $w \to 1$ but remains finite.
 (b) at $T_2=0.15$ where $\mu(T_2) \sim 2$  for $N=50,100,200,400$ :
the distribution $P_1(w)$ vanishes linearly as $w \to 1$.  }
\label{figp1wt1t2}
\end{figure}

The probability distribution $P_1(w)$ of the largest weight seen by a given monomer
is shown on Fig. \ref{figp1w} for low and high temperatures.

At low temperature (Fig. \ref{figp1w} a), this distribution presents a divergent singularity near $w \to 1$
\begin{equation}
P_1(w) \oppropto_{w \to 1} (1-w)^{\mu(T)-1}
\label{p1singw1}
\end{equation}
with a temperature dependent exponent $\mu(T)$, called $\mu$ in analogy with the case
of L\'evy sum of index $\mu$, and with the random energy model where $\mu(T)=T/T_g$ (see the Appendix).
However here in RNA, the pairings free-energies are not independent variables and are
not drawn with the same distribution (as a consequence of the distances involved),
so the full distribution $P_1(w)$ cannot coincide with the L\'evy sums result.
Nevertheless, we find that $P_1(w)$ presents the characteristic Derrida-Flyvbjerg 
singularities at $w=1/n$ (see Appendix ).
The stronger singularity occurs at $w = 1$ and defines the exponent $\mu(T)$ (\ref{p1singw1}),
but the second singularity at $w=1/2$ is also clearly visible on Fig. \ref{figp1w}. 
This shows that for each base in the frozen phase,
all the weight is concentrated on a few pairing partners $j$.

At sufficiently high temperature (Fig. \ref{figp1w} b), 
the distribution $P_1(w)$ does not reach $w=1$ anymore, but vanish
at some maximal value $w_0(T)<1$ 
\begin{equation}
P_1(w) \oppropto_{w \to w_0(T)} (w_0(T)-w)^{\sigma}
\label{p1gapw0}
\end{equation}
with some exponent $\sigma$, that within our numerical precision, 
does not depend on temperature (see below).

So the present results for $P_1(w)$ reveal the importance of two temperatures $T_1 < T_{gap}$
defined as follows. The temperature $T_1$ is defined by 
\begin{equation}
\mu(T_1)=1
\label{deft1}
\end{equation}
in Eq. \ref{p1singw1} : for $T<T_1$, the probability distribution
$P_1(w)$ is divergent as $w \to 1$
(as on Fig. \ref{figp1w} a) ,
whereas for $T>T_1$,  the probability distribution $P_1(w)$ vanishes at $w
= 1$ (see Fig. \ref{figp1wt1t2} b ) . Exactly at $T_1$, $P_1(w=1)$
remains finite (see Fig. \ref{figp1wt1t2} a ).
The second temperature $T_{gap}$ is defined as the last temperature where $P_1(w)$ reaches $w=1$
with some exponent $\mu(T_{gap})$. For $T>T_{gap}$, a gap appears in (Eq. \ref{p1gapw0})

\begin{eqnarray}
w_0(T_{gap}) && =1 \\
w_0(T_{gap} +\epsilon) && <1
\end{eqnarray}
We find that $T_{gap}$ is clearly above $T_1 \sim 0.095$ since 
at $T_2 \sim 0.15$, $P_1(w)$ still reaches $w=1$ with a finite slope
corresponding
to $\mu(T_2) \sim  2$ as shown on Fig. \ref{figp1wt1t2} b.

\subsection{Probability distribution $P_2(w)$
 of the second largest weight seen by a given monomer}

\begin{figure}[htbp]
\includegraphics[height=6cm]{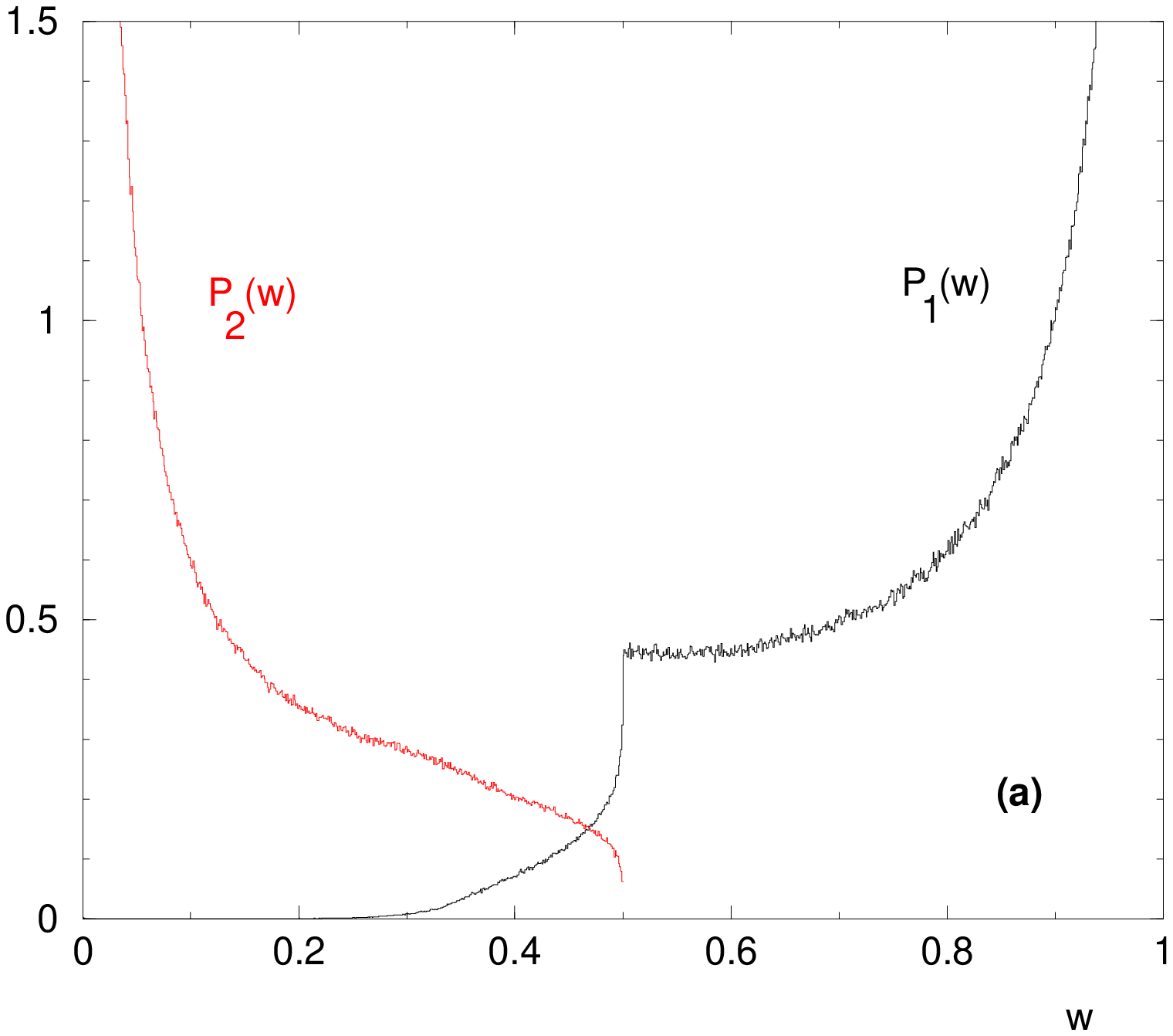}
\hspace{1cm}
\includegraphics[height=6cm]{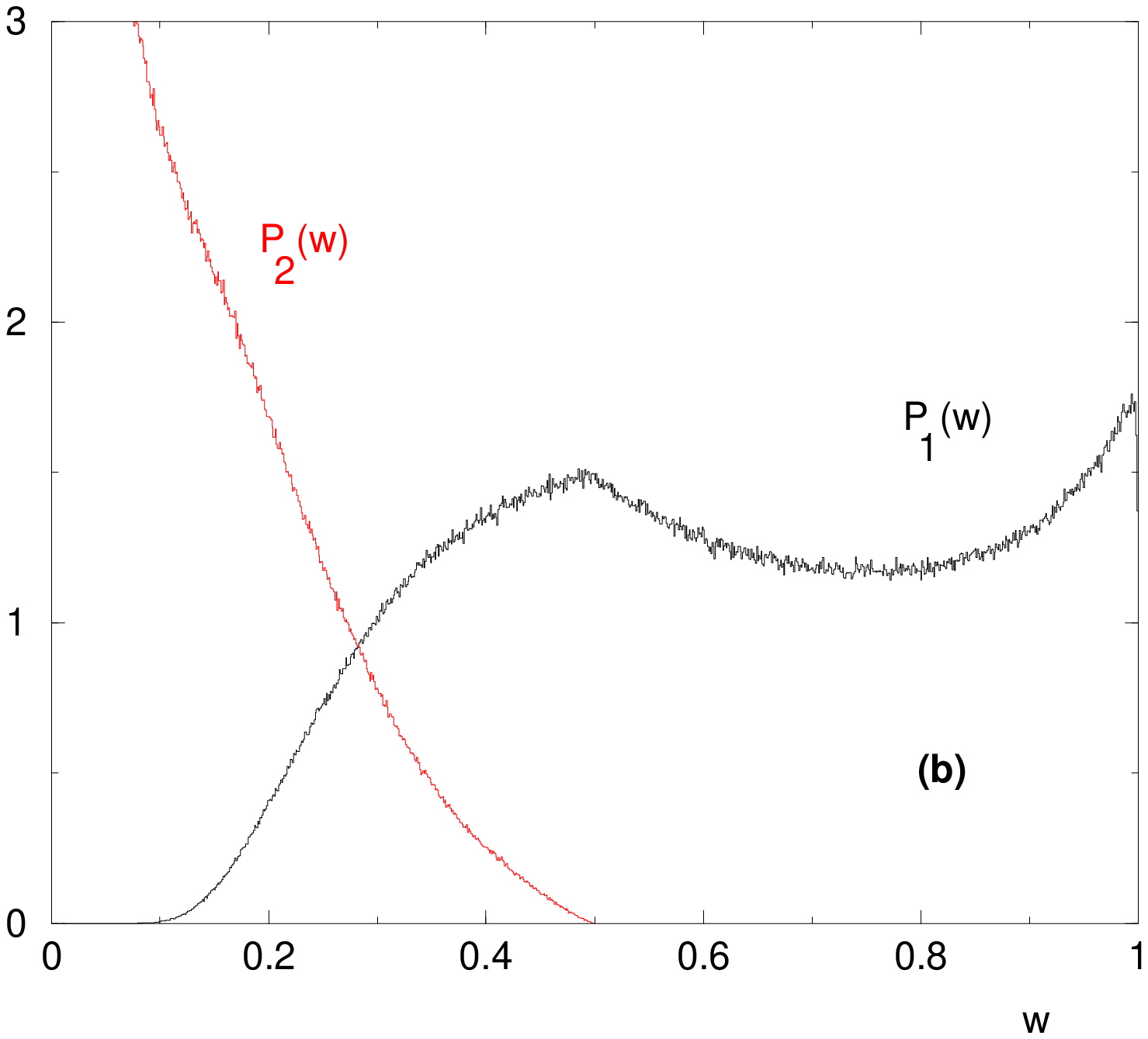}
\caption{(Color online)
  Probability distribution $P_1(w)$ and $P_2(w)$
 of the two largest weights seen by a given monomer (see Eq \ref{wmax})
(a) for $T=0.02$ (here $N=200$) there exists an infinite slope
for $P_2(w)$ and $P_1(w)$ as $w \to (1/2)^-$
(b) for $T_1 \sim 0.095$ (here $N=400$), there exists a cusp
for $P_2(w)$ and $P_1(w)$ as $w \to (1/2)^-$  }
\label{figp2w}
\end{figure}

We show on Fig. \ref{figp2w} the probability distribution $P_2(w)$
of the second largest weight.
The main singularities of $P_2(w)$ are the divergence near $w \to 0$
and the singularity near $w \to (1/2)^-$, which is complementary
to the singularity of $P_1(w)$ at the same point $w \to (1/2)^-$ \cite{Der_Fly}
(see Eq. \ref{intervals} and explanations in Appendix).
For $T=0.02$ (Fig. \ref{figp2w} a), there exists an infinite slope
for $P_2(w)$ and $P_1(w)$ as $w \to (1/2)^-$.
For $T=T_1 \sim 0.095$ (Fig. \ref{figp2w} b), there exists a cusp
for $P_2(w)$ and $P_1(w)$ as $w \to (1/2)^-$.

\subsection{Probability distribution $\Pi(Y_2)$
 of the parameter $Y_2$}

\begin{figure}[htbp]
\includegraphics[height=6cm]{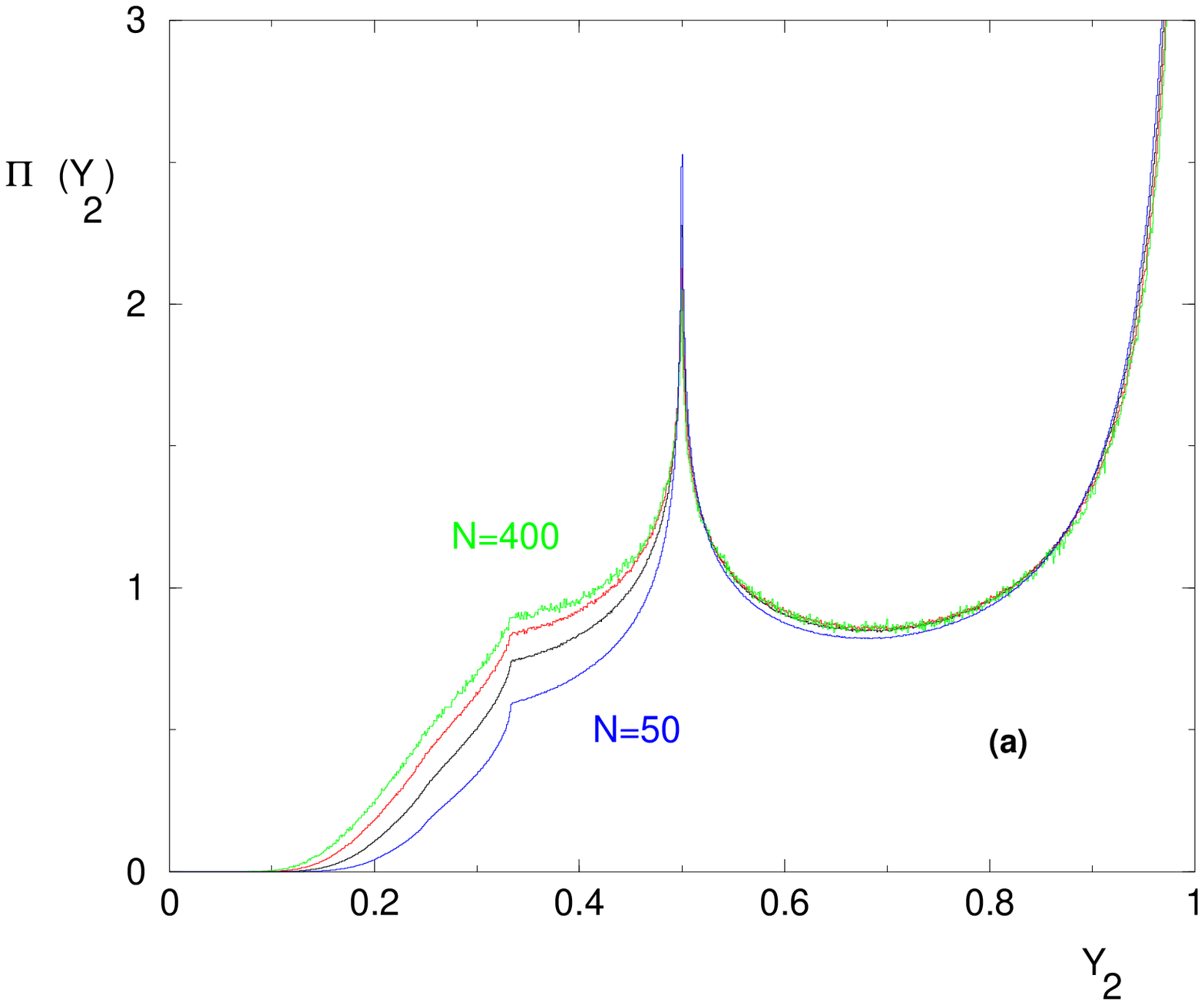}
\hspace{1cm}
\includegraphics[height=6cm]{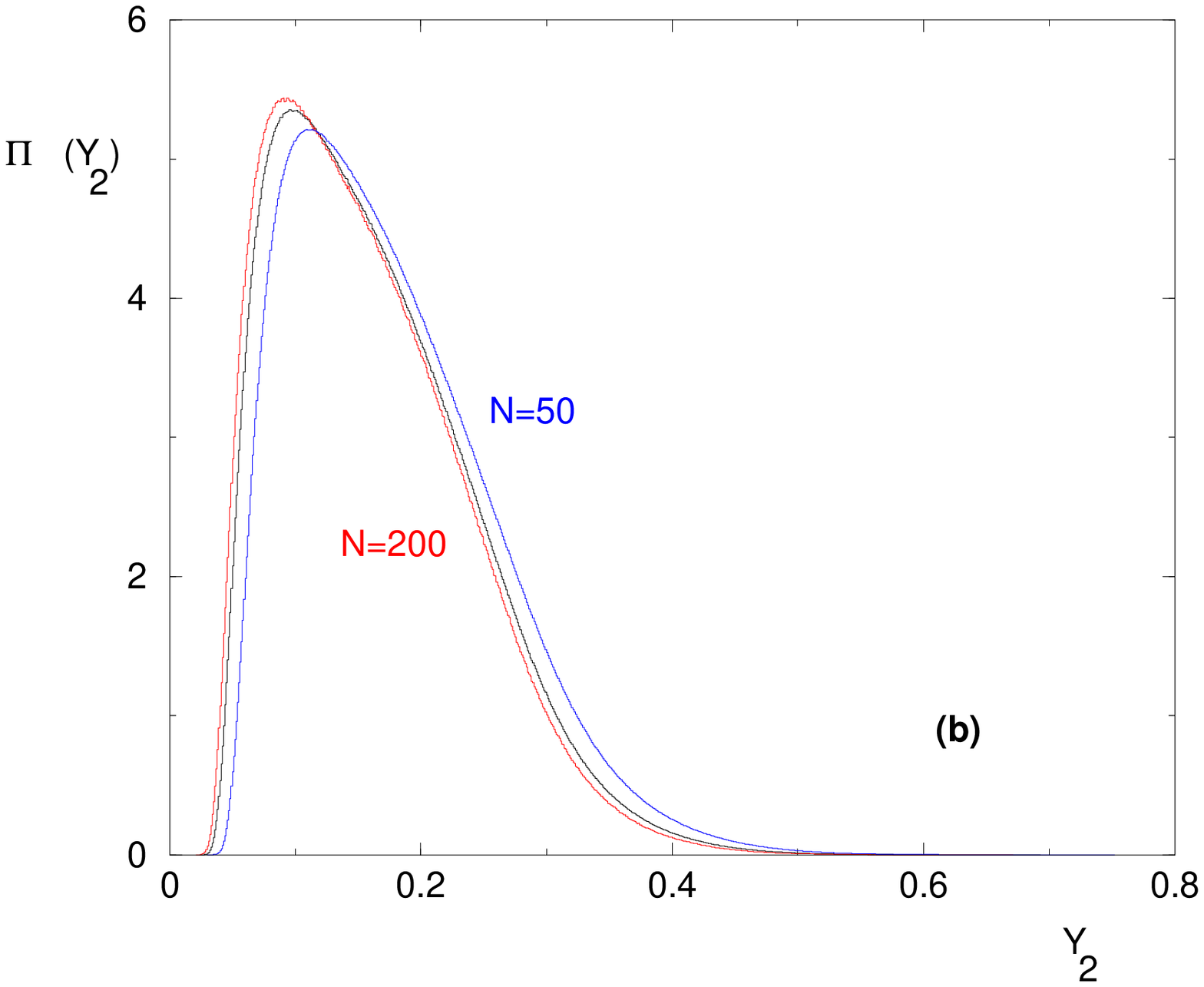}
\caption{(Color online) Probability distribution $\Pi(Y_2)$
 of the parameter $Y_2$ (see Eq \ref{y2rna})
(a) at $T=0.05$ (low-temperature phase) for $N=50,100,200,400$ :
the characteristic Derrida-Flyvbjerg 
singularities at $Y_2=1$, $Y_2=1/2$ and at $Y_2=1/3$ are clearly visible.
(b) at $T=0.4$ (high-temperature phase) for $N=50,100,200$ :   
the distribution $\Pi(Y_2)$ does not reach $Y_2=1$ anymore, but 
presents a gap.}
\label{figpiy2}
\end{figure}

\begin{figure}[htbp]
\includegraphics[height=6cm]{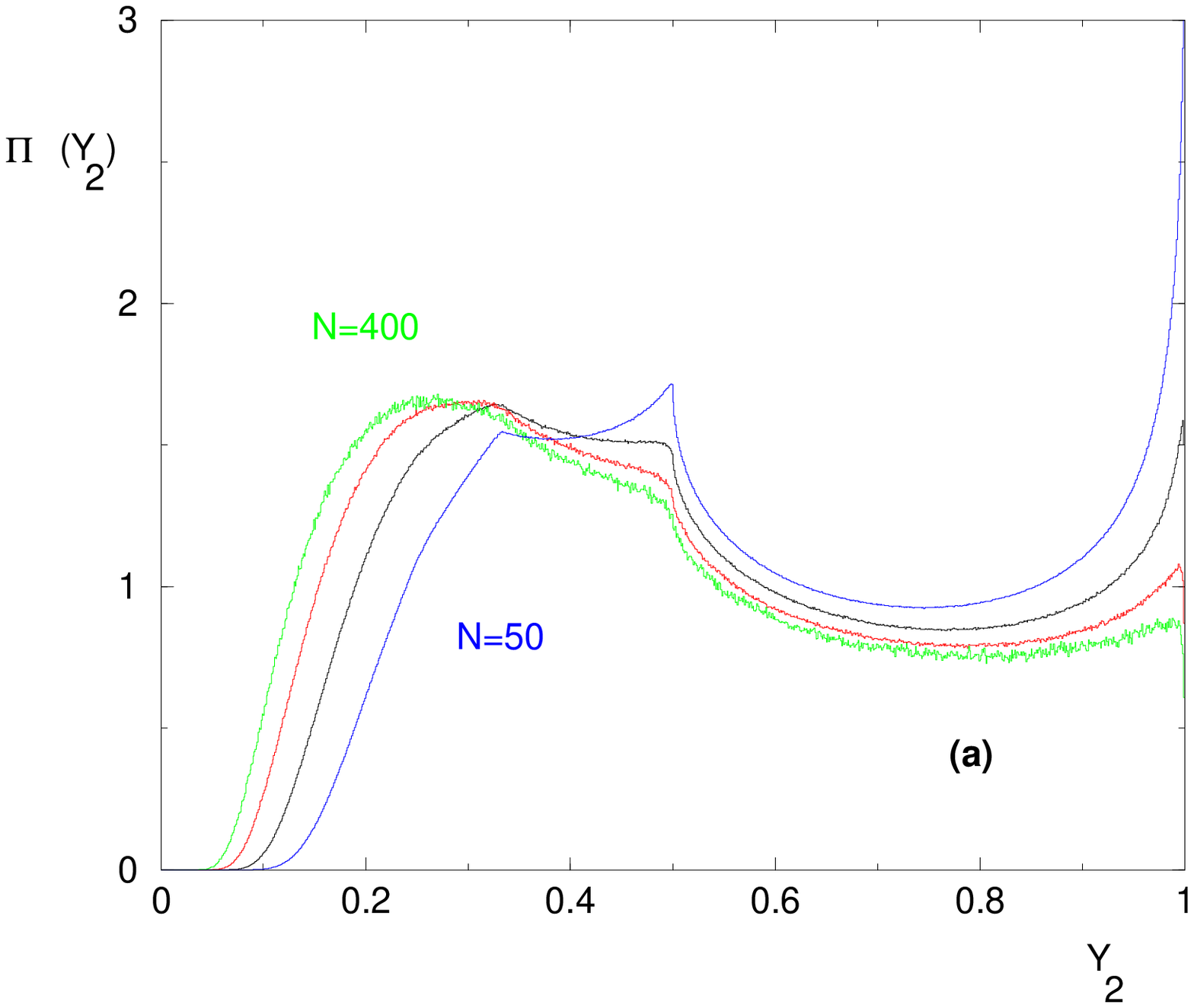}
\hspace{1cm}
\includegraphics[height=6cm]{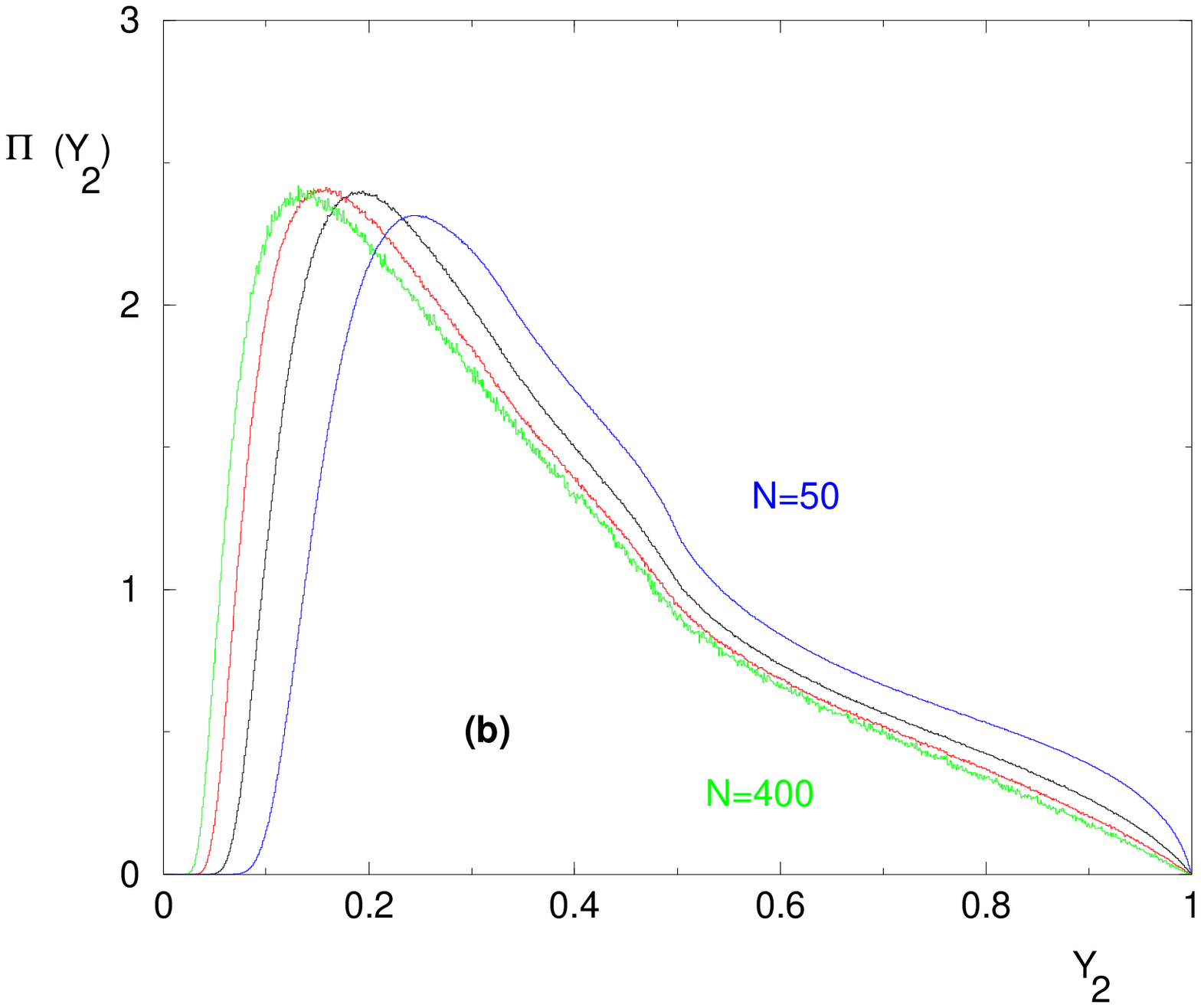}
\caption{(Color online) Probability distribution $\Pi(Y_2)$
 of the parameter $Y_2$ (see Eq \ref{y2rna})
(a) at $T_1 \sim 0.095$ where $\mu(T_1)= 1$ for $N=50,100,200,400$ :
$\Pi(Y_2)$ does not diverge anymore as $Y_2 \to 1$ but remains finite
(b) at $T_2=0.15$   where $\mu(T_2) \sim 2$ for $N=50,100,200,400$ :
the distribution  $\Pi(Y_2)$ vanishes linearly as $Y_2 \to 1$.}
\label{figpiy2t1t2}
\end{figure}

The parameter $Y_2$ defined in Eq. \ref{y2rna} can reach the value $Y_2 \to 1$
only if the maximal weight $w^{max}$ also reaches $w^{max} \to 1$.
As a consequence, the probability distribution $\Pi(Y_2)$
has the same singularity near $Y_2 \to 1$ as in Eq. (\ref{p1singw1})
\begin{equation}
\Pi(Y_2) \oppropto_{Y_2 \to 1} (1-Y_2)^{\mu(T)-1}
\label{pising1}
\end{equation}
for $0 < T < T_{gap}$, (see Fig.\ref{figpiy2} a),
   whereas a gap appear for $T>T_{gap}$, as shown on Fig. \ref{figpiy2} b.

For $T<T_{gap}$,  the distribution
$\Pi(Y_2)$ presents the characteristic Derrida-Flyvbjerg 
singularities at $Y_2=1/n$ (see Appendix ) : on
Fig. \ref{figpiy2} a, beyond the main singularity at $Y_2=1$, the
secondary singularities at $Y_2=1/2$ and at $Y_2=1/3$ are clearly
visible.

The distribution $\Pi(Y_2)$ is shown on Fig. \ref{figpiy2t1t2}
for the two temperatures $T_1$ and $T_2$ corresponding to
$\mu(T_1) = 1$ and $\mu(T_2) \sim 2$, and can be compared with
the distribution $P_1(w)$  on Fig. \ref{figp1wt1t2}.

\subsection{Density $f(w)$ }

\begin{figure}[htbp]
\includegraphics[height=6cm]{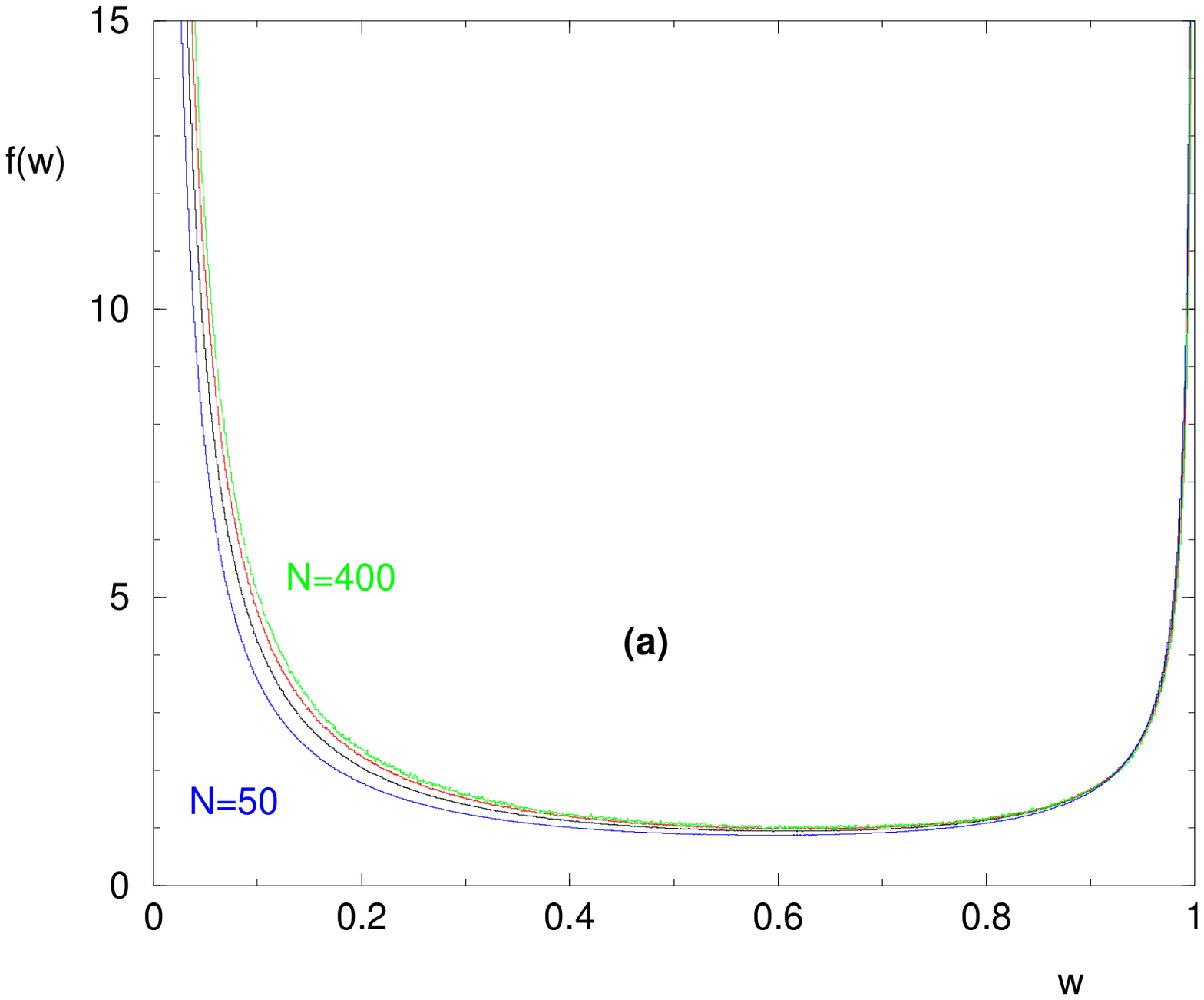}
\hspace{1cm}
\includegraphics[height=6cm]{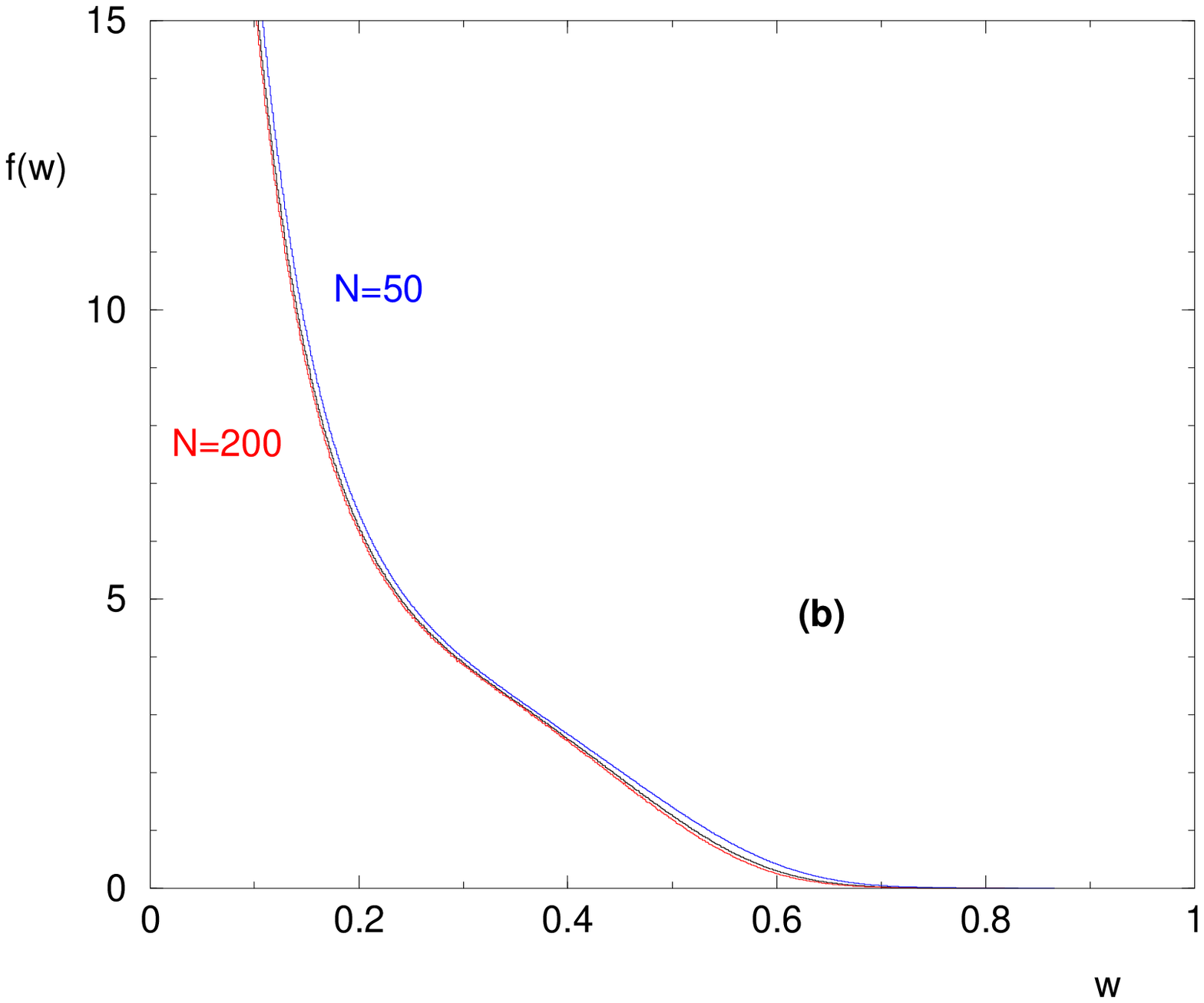}
\caption{(Color online) Density $f(w)$
 of weights seen by a given monomer (see Eq \ref{densityfrna})
(a) at $T=0.05$ (low-temperature phase) for $N=50,100,200,400$.
Near $w \to 1$, $f(w)$ presents the same singularity as $P_1(w)$.
 (b)at $T=0.4$ (high-temperature phase) for $N=50,100,200$ :
$f(w)$ vanishes at some value $w_0(T)<1$. }
\label{figdensityf}
\end{figure}

\begin{figure}[htbp]
\includegraphics[height=6cm]{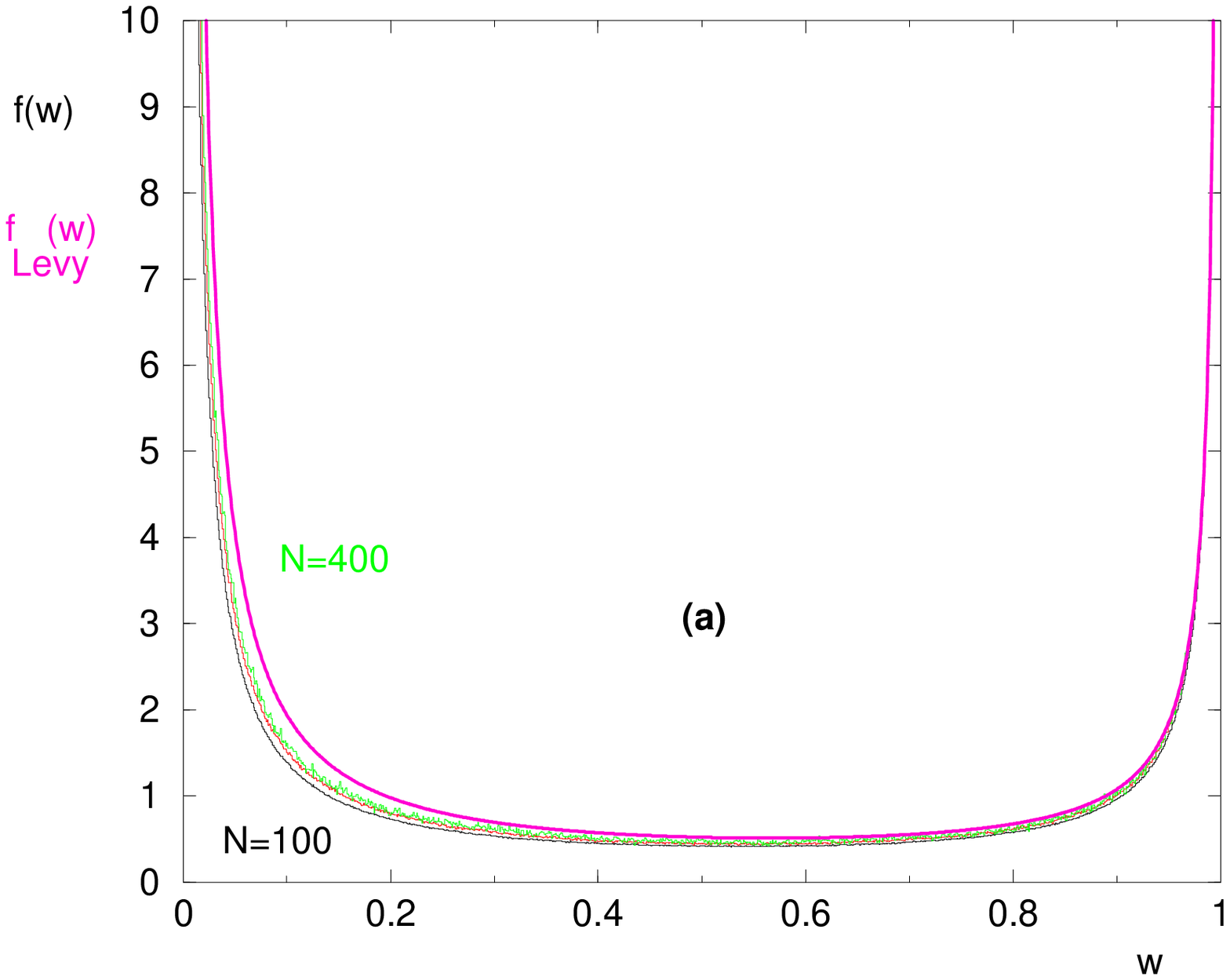}
\hspace{1cm}
\includegraphics[height=6cm]{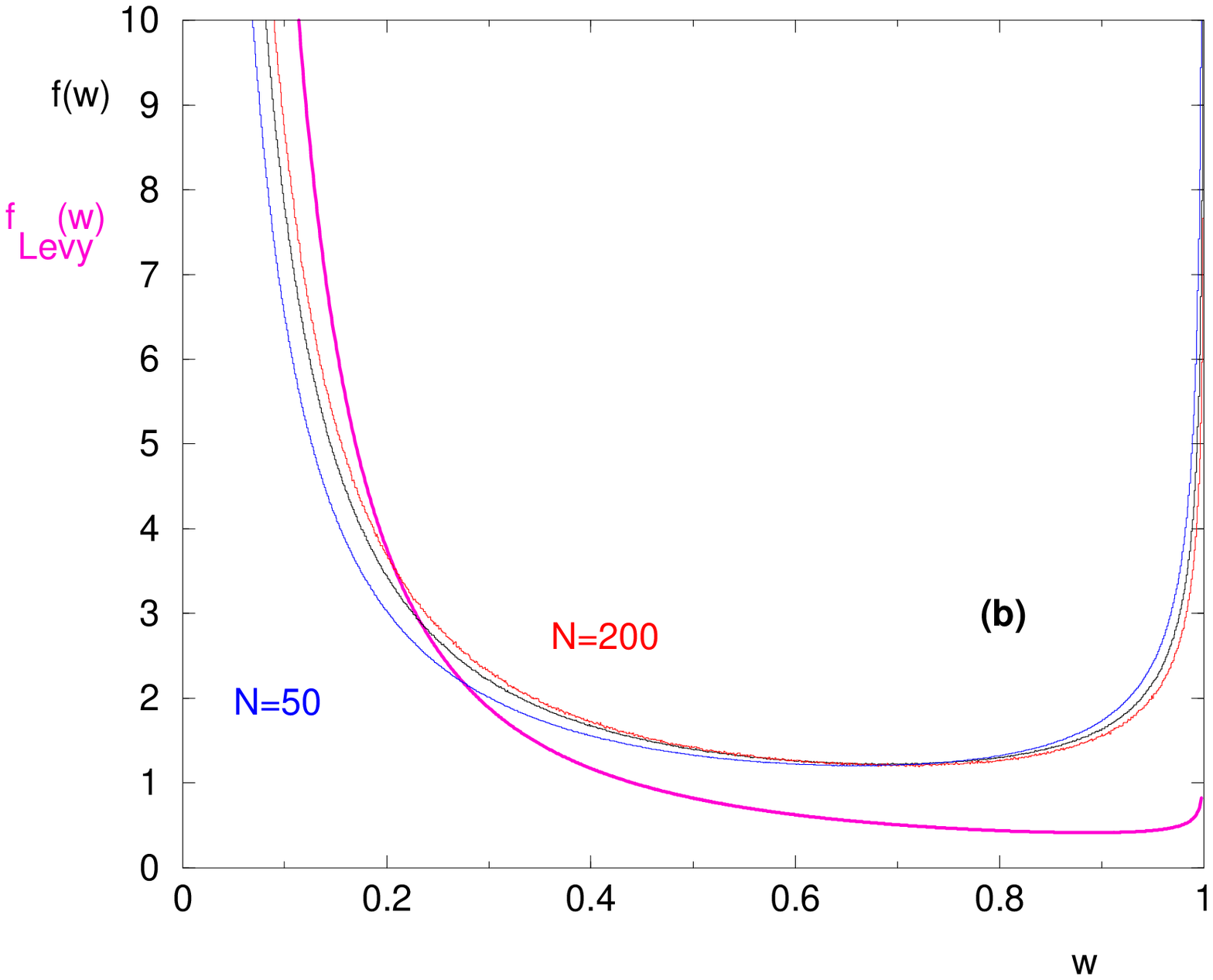}
\caption{(Color online) 
Density $f(w)$
 of weights seen by a given monomer (see Eq \ref{densityfrna})
as compared to $f_{Levy}(w)$ (Equation \ref{densitew}) for the 
corresponding value of $\mu$ (thick curve)
(a) at $T=0.02$ where $\mu (0.02) \sim  0.13 $ ( for $N=100,200,400$ ) :
 $f(w)$ is rather close to the density $f_{Levy}(w)$.
 (b) at $T=0.08$ where $\mu (0.08) \sim  0.77 $ (for $N=50,100,200$) :
there is now a big difference between the density $f(w)$ measured
for RNA and the density $f_{Levy}(w)$.  }
\label{figdensitylevy}
\end{figure}

The density $f(w)$ introduced in Eq. (\ref{densityfrna}) is shown on
Fig. \ref{figdensityf} 
at low and high temperature respectively.
By construction, this density coincides with the maximal weight
distribution $P_1(w)$ for $w>1/2$, with the sum ($P_1(w)+P_2(w)$) of
the two largest weight distributions for $1/3<w<1/2$, and so on
\cite{Der_Fly} (see Eq. \ref{intervals}). 
As a consequence, $f(w)$ has the same singularity near 
$w \to 1$ as $P_1(w)$ (Eq. \ref{p1singw1} ),
and the same gap (Eq. \ref{p1gapw0} ) as long as $w_0(T)>1/2$.
The only other singularity is near $w \to 0$ where $f(w)$ diverges 
in a non-integrable manner, because in the $N \to \infty$, there is an
infinite number of vanishing weights (only the product $(w f(w))$ has
to be integrable at $w=0$ as a consequence of the normalization
condition of Eq. \ref {normay1}).

On Fig. \ref{figdensitylevy}, we compare the density $f(w)$
measured in RNA with the density $f_{Levy}(w)$ (Equation \ref{densitew})
describing the weight statistics in L\'evy sums of independent variables.
For small $\mu \ll 1$, the two density are rather close
(Fig. \ref{figdensitylevy} a). For larger $\mu$,
they are very different, because the density $f_{Levy}(w)$ disappears
at the critical value $\mu_c=1$ (see  the denominator of
Equation \ref{densitew}), whereas for RNA the density $f(w)$ 
 exists beyond $\mu=1$.

\subsection{ Moments  $\overline{Y_k(i)} $ }

\begin{figure}[htbp]
\includegraphics[height=6cm]{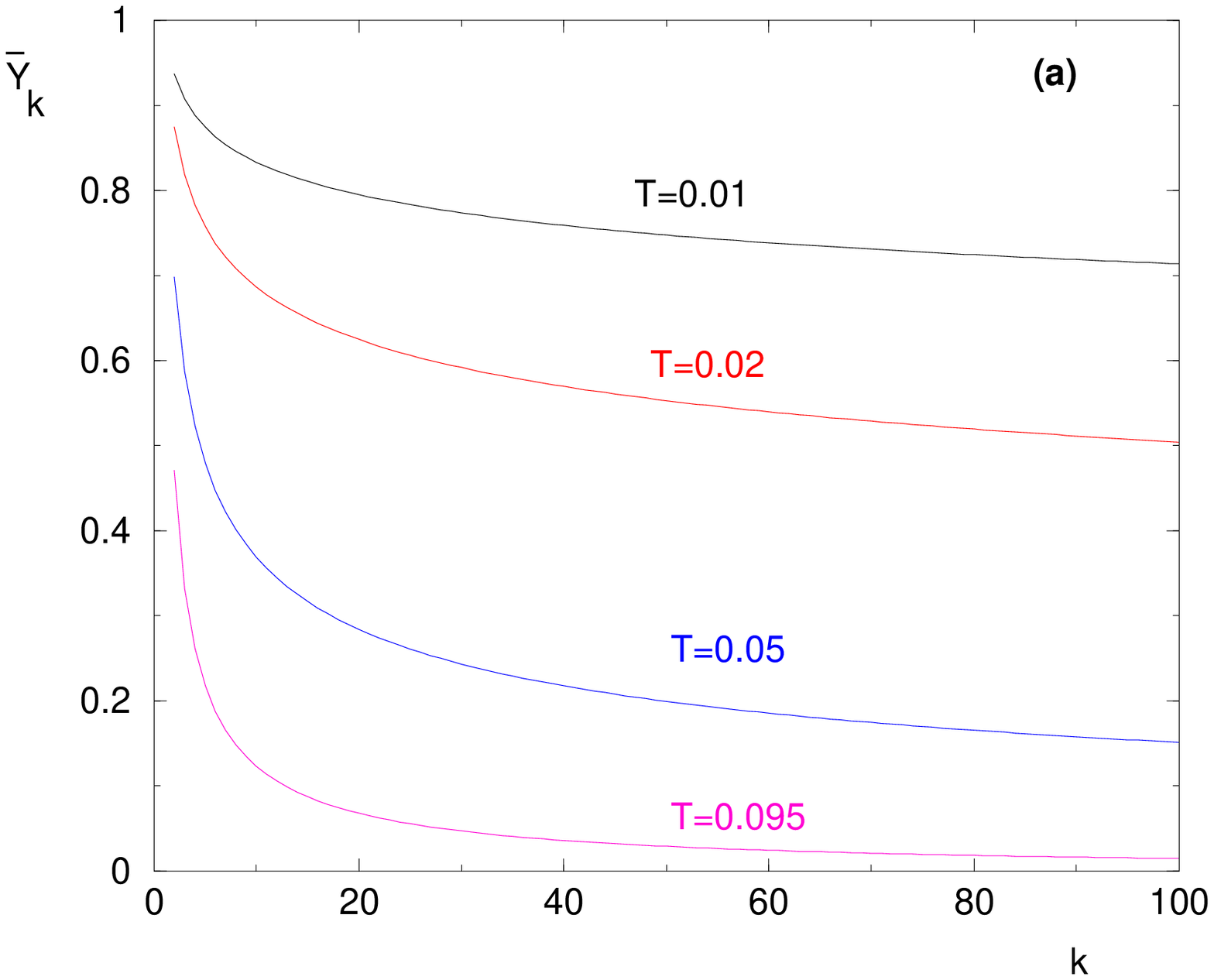}
\hspace{1cm}
\includegraphics[height=6cm]{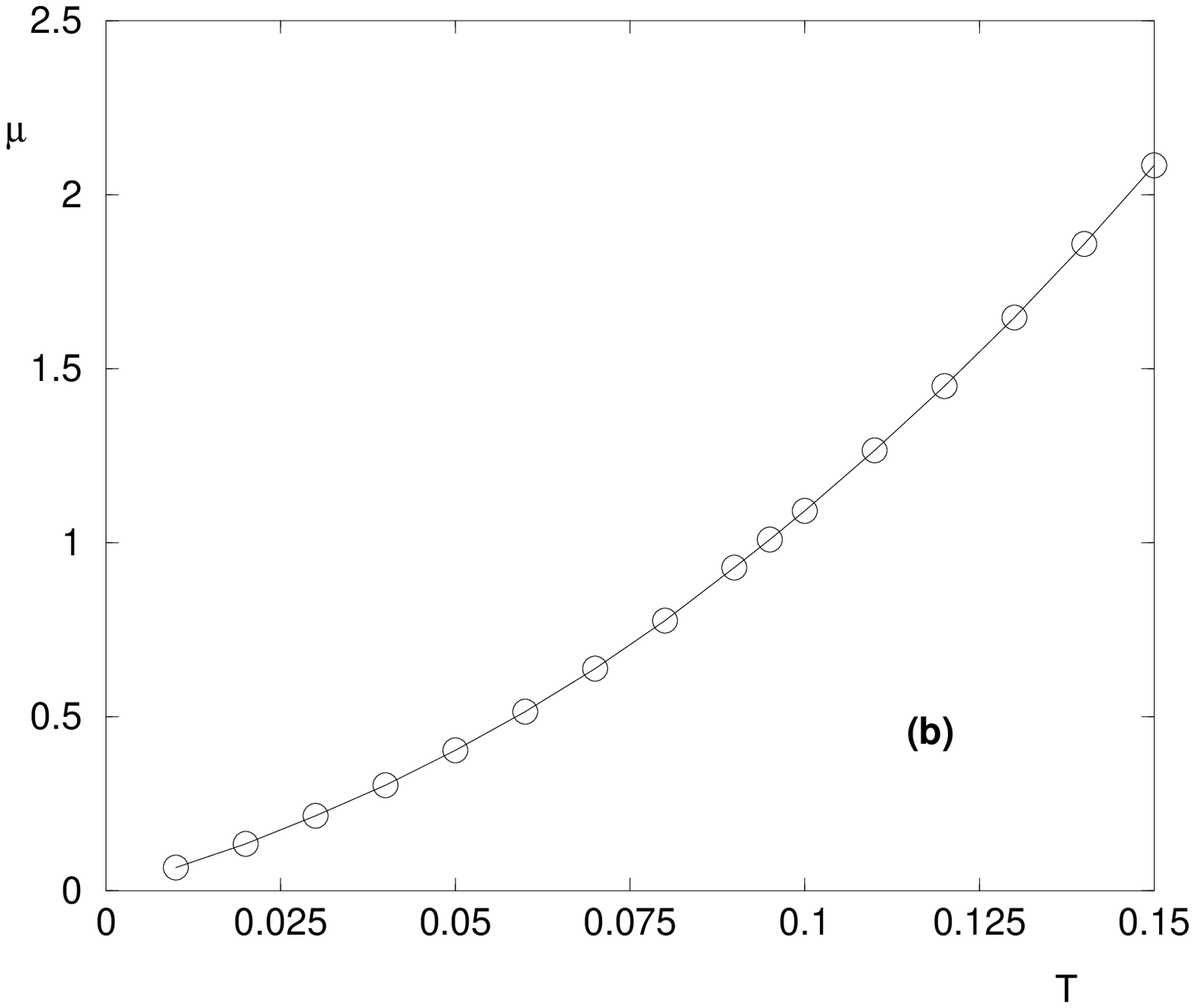}
\caption{(Color online) 
(a) Decay of the moments $\overline{Y_k(i)} $ of Eq. \ref{ykdecay}
as a function of $k \leq 100$ for $N=800$ and
 $T=0.01,0.02,0.05,0.095$
(b)
Exponent $\mu(T)$ as measured from the slope of the log-log decay
in the asymptotic region.}
\label{figmutemp}
\end{figure}

The moments $\overline{Y_k(i)} $ (Eq. \ref{ykrna}) for various temperature are shown on Fig. \ref{figmutemp} 
as functions of $k \leq 100$. The decay for large $k$ directly reflects the behavior of
the distribution of the density $f(w)$ near its maximal value, as can be seen on Eq. \ref{ykf}.
For $0<T\leq T_{gap}$, where $P_1(w)$ and $f(w)$ behaves near $w \to 1$ as in Eq. \ref{p1singw1},
the decay in $k$ follow a power-law of exponent $\mu(T)$
\begin{equation}
\overline{Y_k(i)} \oppropto_{k \to \infty}  \frac{1}{k^{\mu(T)}} 
 \  \  {\rm for } \ \ T \leq T_{gap}
\label{ykdecay}
\end{equation}
For $T>T_{gap}$, where there exists a gap $w_0(T)$ for $P_1(w)$,  Eq. \ref{p1gapw0}
also applies to $f(w)$ as long as $w_0(T)>1/2$ (since $f(w)=P_1(w)$ for $w>1/2$ as mentioned above).
This implies an exponential decay
\begin{equation}
\overline{Y_k(i)} \oppropto_{k \to \infty}  \frac{ (w_0(T))^k}{k^{1+\sigma}}  \  \  {\rm for } \ \ T > T_{gap}
\label{ykgap}
\end{equation}
Numerically, the measure of the decay of $\overline{Y_k(i)}$ is the most convenient way
to localize the temperature $T_{gap}$ where the gap appears, and to
measure the exponents.The temperature $T_{gap}$ where the gap appears is
found to be
\begin{equation}
T_{gap} \sim T_2 \sim 0.15 
\end{equation}
The exponent $\sigma$
 seems to be independent of $T$
\begin{equation}
\sigma \sim 0.5 
\end{equation}
The exponent $\mu(T)$ grows with the temperature for $0<T \leq T_{gap}$,
as shown on  Fig. \ref{figmutemp} b :
the temperature $T_1$ where $\mu(T_1)=1$ is
\begin{equation}
T_1 \sim 0.095
\end{equation}

In contrast with the REM (see Appendix) where $\mu(T)=T/T_g$ is linear
in the whole low temperature phase, Fig \ref{figmutemp} b  presents
some curvature, which probably reflects the presence of some
entropy. However, in the limit of very low temperature, the exponent
$\mu(T)$ is linear in $T$ and the coefficient depends on the droplet
density as we now explain.  

\subsection{ Droplet analysis at order $T$ in temperature }

At $T=0$, there exists a unique ground-state where each monomer $(i)$
is paired with its ground-state partner $j_{gs}(i)$.
In the droplet analysis of disordered systems \cite{Fis_Hus_SG,Fis_Hus,twolevel},
various observables
can be computed as first order in $T$ in terms of the density
$\rho(E=0,\lambda)dE$ of two-level excitations of energy $E \to 0$
and size $\lambda$. For instance the specific heat and the overlap 
are given in 1D disordered spin chains by \cite{twolevel,chenma}
\begin{eqnarray}
C(T) && \opsimeq_{T \to 0} T \frac{\pi^2}{6} \int d \lambda
\rho(E=0,\lambda) +O(T^2)\\
1-q_{EA}(T) && \opsimeq_{T \to 0} 2 T \int d \lambda \  \lambda
\rho(E=0,\lambda)+O(T^2)
\end{eqnarray}
i.e. the specific heat is related to the number of excitations
 whereas the overlap involves the number of spins belonging to
excitations.
We may apply this droplet analysis to $\overline{Y_k(i)}$ :
the contribution at order $T$ comes from the monomers $i$
belonging to a droplet of energy $E \to 0$ :
the pair with the ground-state partner has for weight $1/(1+e^{-\beta E})$,
whereas the pair with the droplet partner has the complementary weight
$e^{-\beta E}/(1+e^{-\beta E})$. Within this two-level description,
one gets
\begin{eqnarray}
1- \overline{Y_k(i)} && \opsimeq_{T \to 0}  \int dE 
\int d\lambda \lambda \rho(E,\lambda)
\left[ 1 - \left( \frac{1}{1+e^{-\beta E}} \right)^k
- \left( \frac{e^{-\beta E}}{1+e^{-\beta E}} \right)^k
 \right] \nonumber \\
&& \opsimeq_{T \to 0}  T \left( \int d \lambda \lambda
 \rho(E=0,\lambda) \right) I_k +O(T^2)
\label{droplet}
\end{eqnarray}
where the integral $I_k$ 
\begin{equation}
I_k= \int_{1/2}^{1} \frac{dp}{p (1-p)} \left[ 1-p^k -(1-p)^k \right]
= \sum_{m=1}^{k-1} \frac{1}{m}
\end{equation}
behaves logarithmically at large $k$
\begin{equation}
I_k \opsimeq_{k \to \infty} \ln k
\end{equation}
The comparison of this droplet analysis with Eq.\ref{ykdecay}
indicates that the exponent $\mu(T)$ should increase from $\mu(T=0)=0$
linearly in $T$ with a coefficient related to the droplet density
\begin{equation}
\mu(T) \simeq  T \left( \int d \lambda  \lambda
\rho(E=0,\lambda) \right) +O(T^2) 
\end{equation}

\section{  Study of spatial properties }

\label{spatial} 

In the last Section, we have studied in details the statistics of the weights
independently of the distance and identified two important temperatures $T_1$
and $T_{gap}$. We now turn to the analysis various spatial properties
to clarify the meaning of $T_1$ and $T_{gap}$ for the pair length statistics.

\subsection{  Weight statistics for long-range pairs }

The density $f(w)$ defined in Eq. \ref{densityfrna} can be decomposed 
into $l$-dependent components as
\begin{equation}
f(w) = \sum_{l} f_l(w)
\end{equation}
where $f_l(w)$ represents the density of weight of pairs of length $l$.
At $T=\infty$, these densities are concentrated
on a single $l$-dependent value (see Eq. \ref{pairijtinfty}
 for $N \to \infty$)
\begin{equation}
\left[f_l(w) \right]_{T=\infty} \propto \delta(w - \frac{a}{l^{3/2}})
\label{histowtinfty}
\end{equation}
whereas at zero temperature (see Eq. \ref{pijt0}  for $N \to \infty$),
a weight is either $0$ (if the pair is not in the ground state)
 or $1$ (if the pair belongs to the ground state)
\begin{equation}
\left[f_l(w) \right]_{T=0} \sim b_l \delta(w)+
c_l \delta(w-1)
\label{histowt0}
\end{equation}
where the amplitude $c_l$ of the existing weights
(see Eq. \ref{pijt0}  for $N \to \infty$)
decay with $l$ as
\begin{equation}
c_l \propto \frac{1}{l^{\eta(T=0)}} = \frac{1}{l^{1.33}}
\label{coefw1t0}
\end{equation}

\begin{figure}[htbp]
\includegraphics[height=6cm]{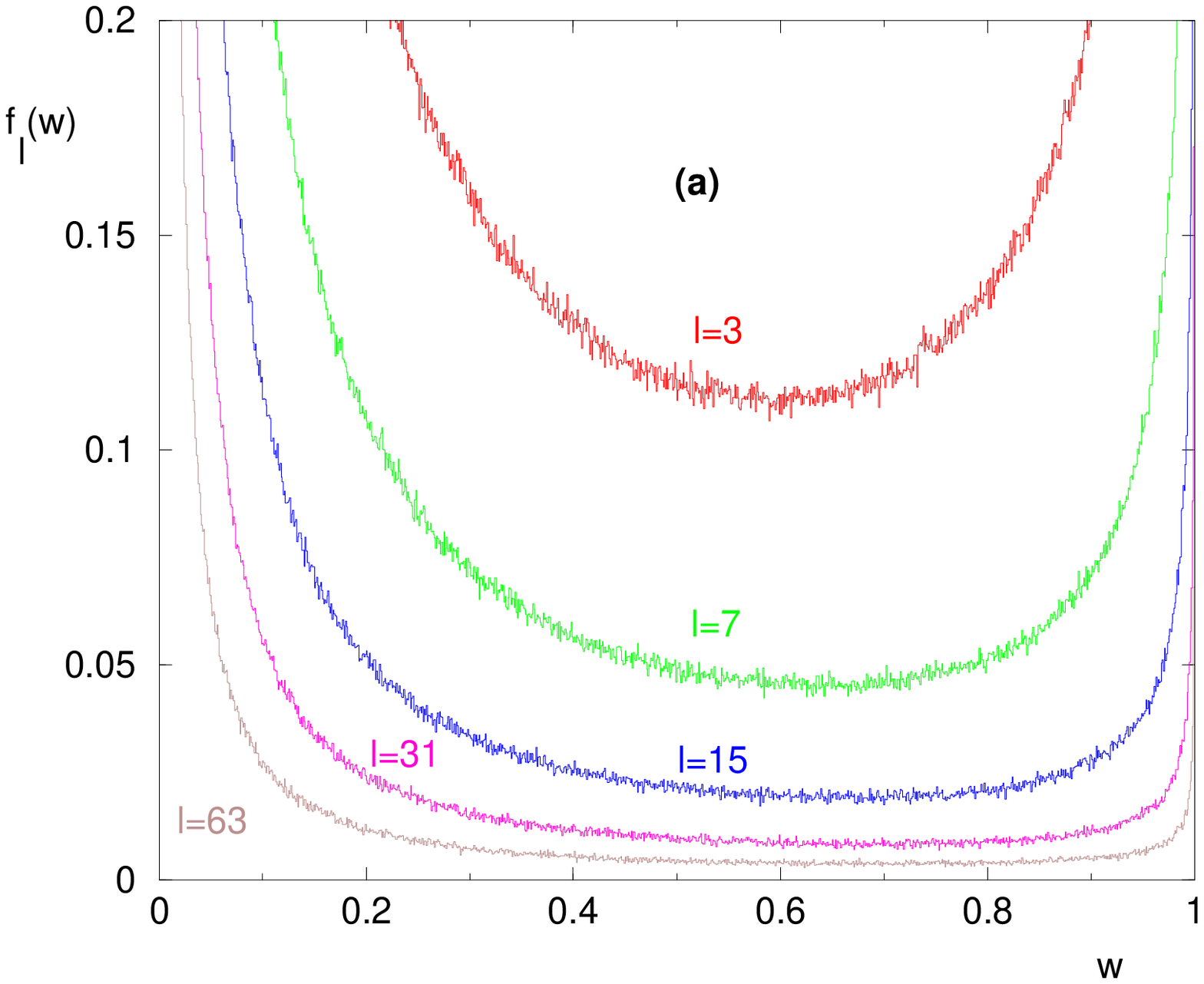}
\hspace{1cm}
\includegraphics[height=6cm]{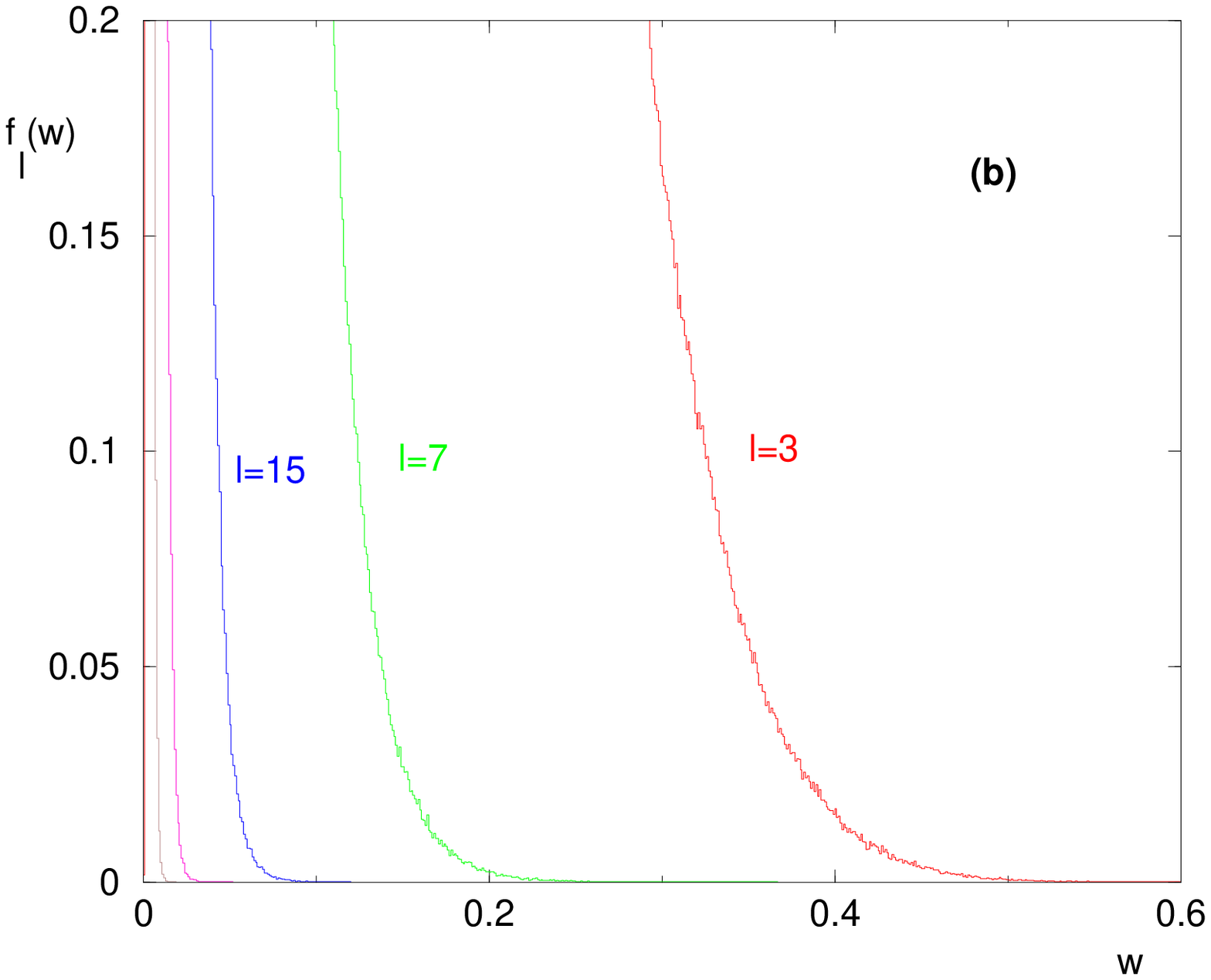}
\caption{(Color online) 
Density $f_l(w)$ of weights for pair lengths $l=3,7,15,31,63$
for size $N=200$
(a) at $T=0.05$, all curves diverge as $w \to 1$.
(b) at $T=0.4$, all curves display an $l$-dependent gap.   }
\label{fighistow1}
\end{figure}

\begin{figure}[htbp]
\includegraphics[height=6cm]{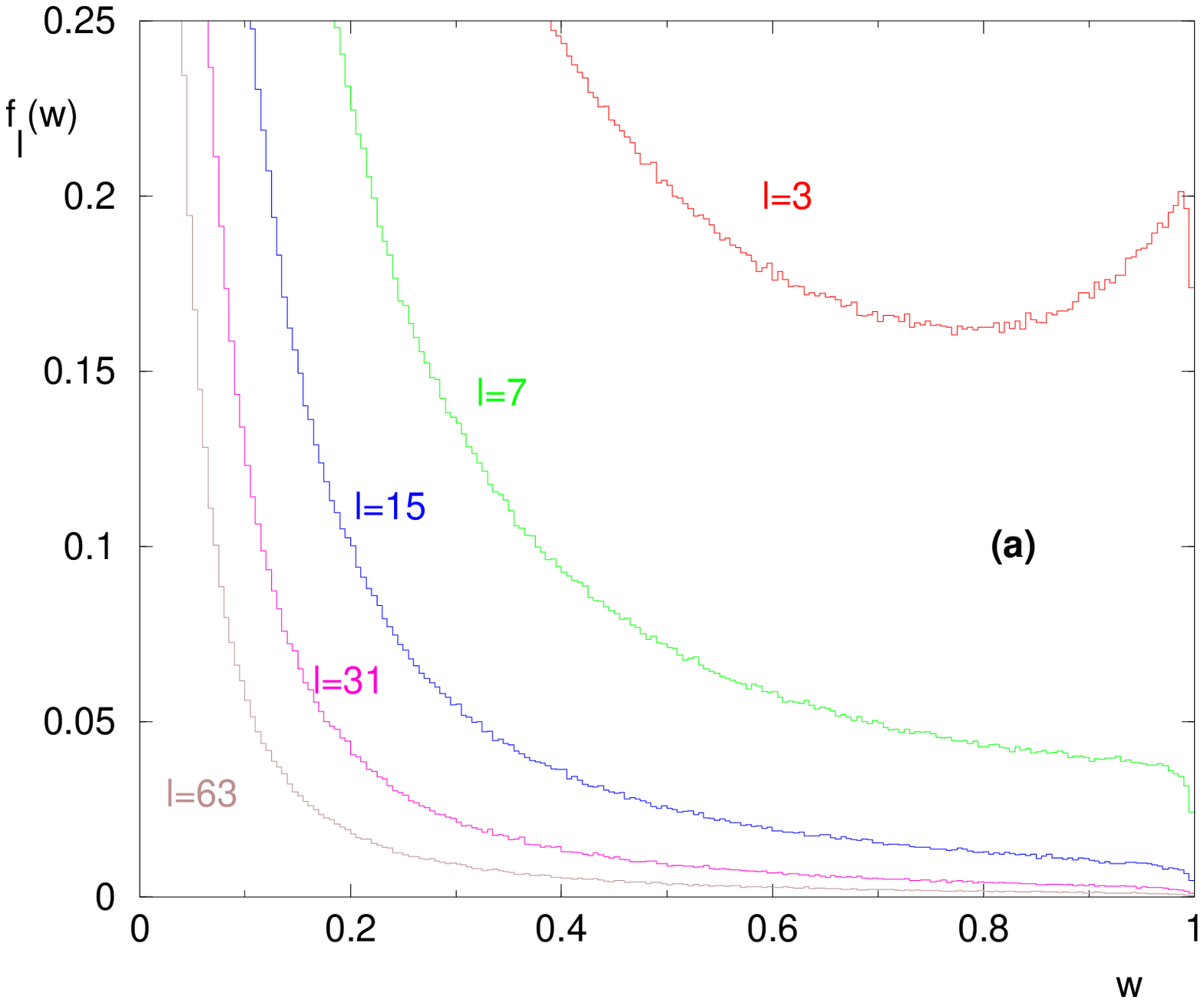}
\hspace{1cm}
\includegraphics[height=6cm]{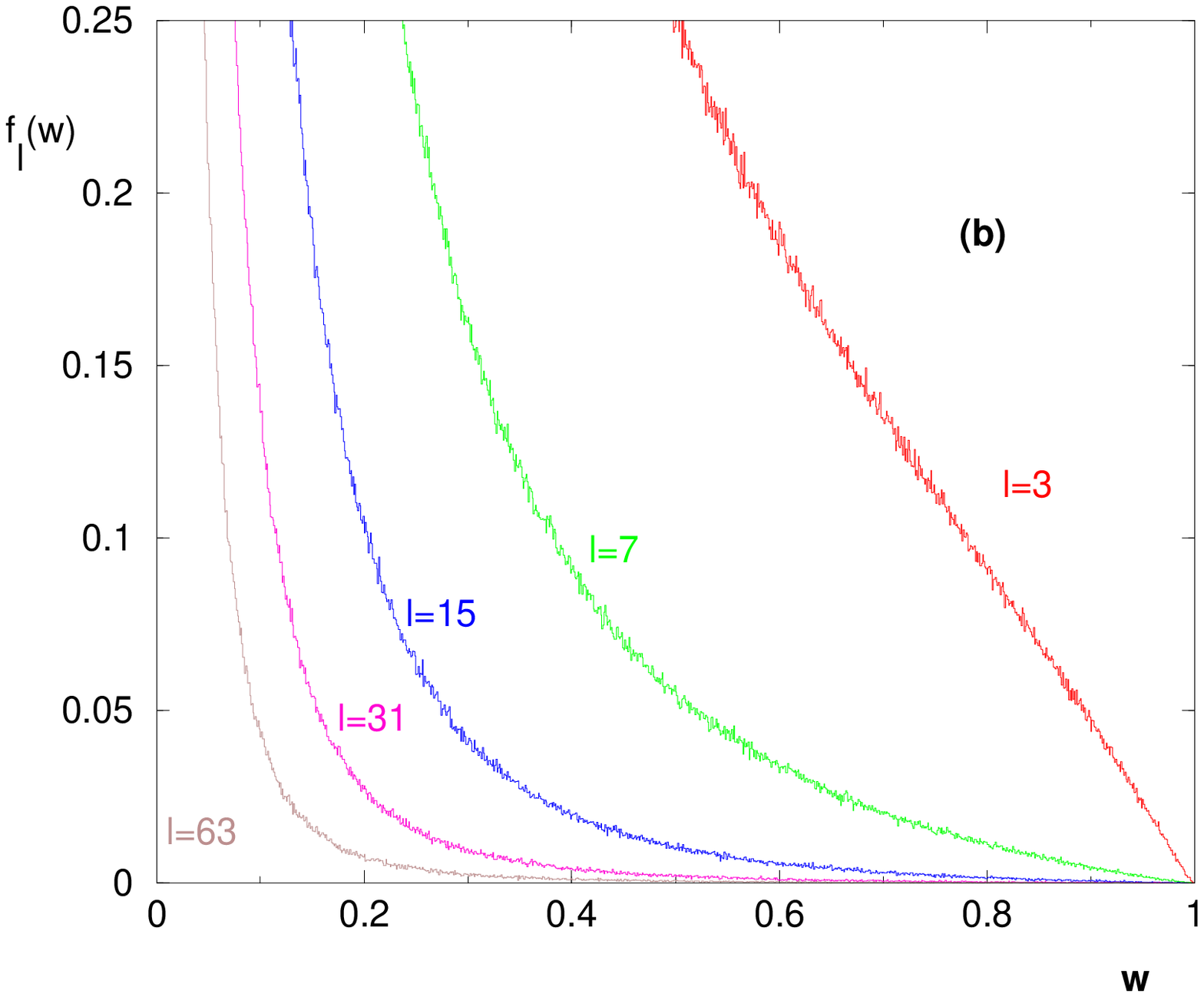}
\caption{(Color online) 
Densities $f_l(w)$ of weights for pair lengths $l=3,7,15,31,63$ 
for size $N=200$
(a) at $T_1 \sim 0.095$, all densities $f_l(w)$ still reach the value $w=1$
(b) at $T_{gap}=0.15$, the densities $f_l(w)$ of small sizes $l$
still reach the value $w=1$, whereas the densities $f_l(w)$ 
of large sizes $l$ don't. }
\label{fighistow2}
\end{figure}

The densities $f_l(w)$ are plotted for various lengths $l$
at low and high temperature respectively on Fig. \ref{fighistow1}.
At $T=0.05$ (Fig. \ref{fighistow1} a), all curves present
divergences at $w \to 0$ and at $w \to 1$, in continuity with
the two delta peaks present at $T=0$ (Eq. \ref{histowt0}).
At $T=0.4$ (Fig. \ref{fighistow1} b)
all curves display an $l$-dependent gap, in similarity with the $l$-dependent
delta peak of the $T=\infty$ limit (Eq. \ref{histowtinfty}).

We show on Fig. \ref{fighistow2} the case of the two important temperatures
$T_1$ and $T_{gap}$.
 At $T_1 \sim 0.095$ (Fig. \ref{fighistow2} a) ,
 all densities $f_l(w)$ still reach the value $w=1$.
At $T_{gap} \sim 0.15$, the densities $f_l(w)$ of small sizes $l$
still reach the value $w=1$, whereas the densities $f_l(w)$ 
of large sizes $l$ don't.

These curves suggest the following picture : \\
(i) for $T<T_1$, all densities $f_l(w)$ diverge near $w \to 1$,
so there exist frozen pairs of all sizes. \\
(ii) for $T_1<T<T_{gap}$, there exist frozen pairs, but only of finite size. \\
(iii) for $T>T_{gap}$, even short pairs are not frozen anymore.
 
We now present various quantitative studies that confirm this scenario.

\subsection{ Statistics of the distance $l_{pref}$ to the preferred partner }

We now consider 
 the probability distribution $P_N^{pref}(l_{pref})$ of the distance
$l_{pref}= \vert j_{pref}(i) -i \vert$ between a base $i$ and its preferred
partner $j_{pref}(i)$, i.e. the monomer $j_{pref} \neq i$ having
the maximal weight (Eq. \ref{wmax}).
At $T=0$, this distribution coincides with the pair distribution of the
ground state (Eq. \ref{pijt0})
\begin{equation}
\left[  P_{N}^{pref}(l) \right]_{T=0} = \overline{ P_{i,i+l} (T=0) }  \sim  \frac{1}{N^{\eta(T=0)}}
 \Phi \left( \frac{l}{N} \right) 
\label{preflt0}
\end{equation}
whereas at $T=\infty$, the maximal weight corresponds to 
the nearest neighbors with $l=1$ for entropic reasons 
(Eqs \ref{pairijtinfty} and \ref{pl1tinfty})
\begin{equation}
\left[ P_{N}^{pref}(l) \right]_{T=\infty} = \delta_{l,1}
\label{prefltinfty}
\end{equation}

\begin{figure}[htbp]
\includegraphics[height=6cm]{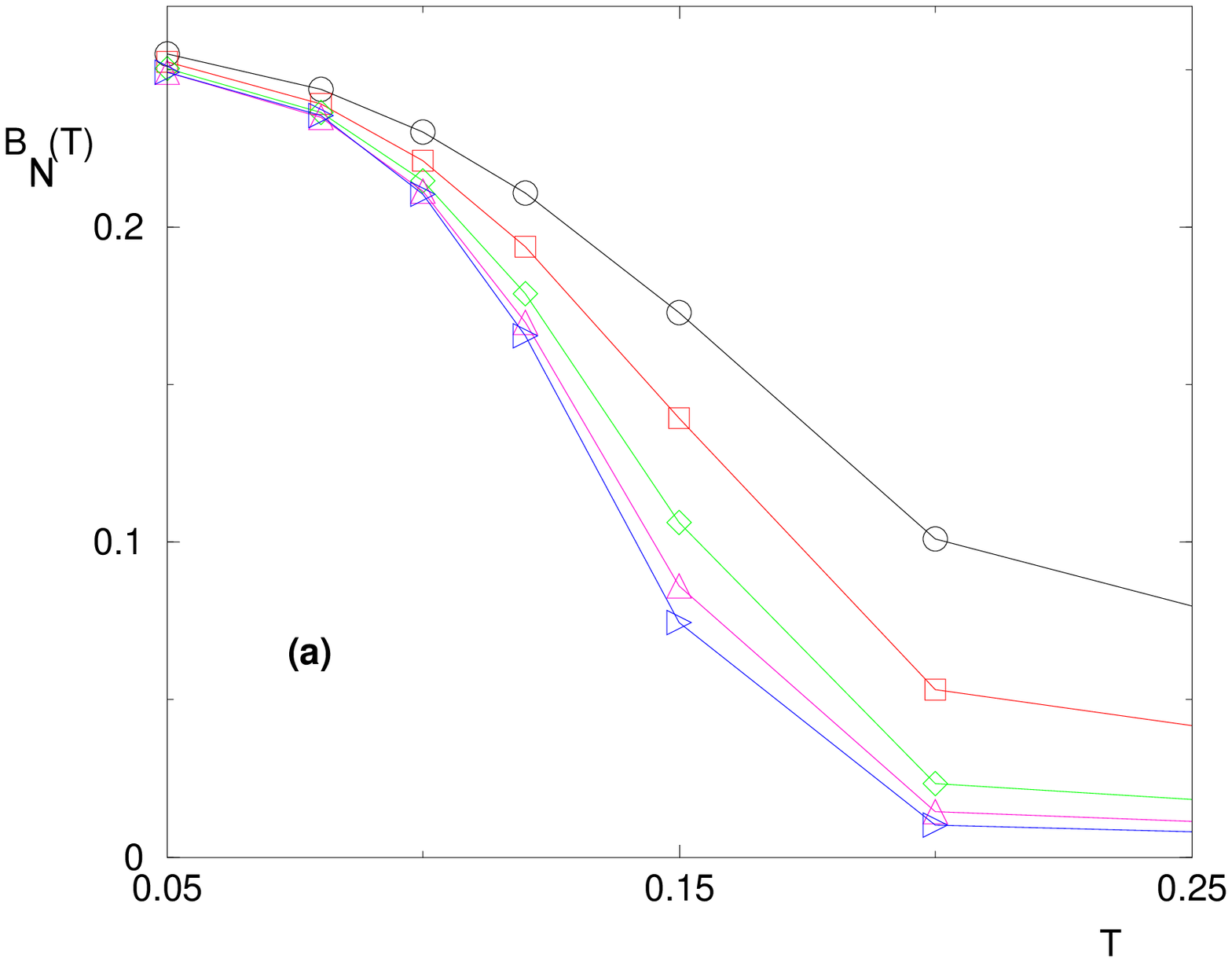}
\hspace{1cm}
\includegraphics[height=6cm]{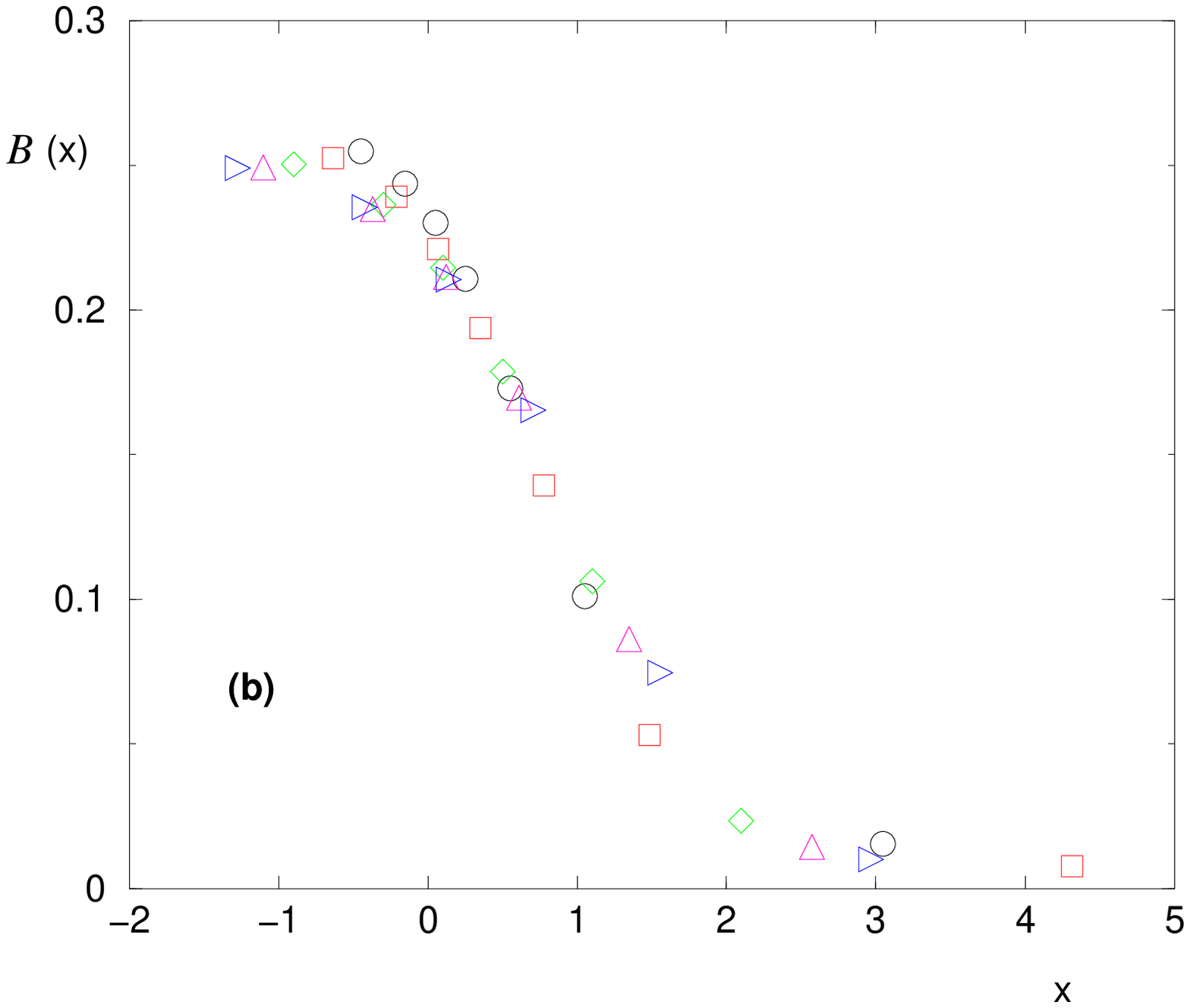}
\caption{(Color online) Ratio $B_N(T)$ defined in  Eq. \ref{binder} :
(a) as a function of temperature $T$ for sizes $N=100$
$(\bigcirc)$, $N=200$ $(\square)$, $N=400$ $(\lozenge)$, $N=600$
$(\vartriangle)$, $N=800$ $(\rhd)$.
(b) finite size scaling of the same data in terms of the variable
$x=(T-T_c) N^{1/{\nu}}$
 (see  Eq. \ref{binderfss}) with $T_c =0.095$ and $\nu=2$ }
\label{figbinder}
\end{figure}

So the first moment of this distribution
\begin{equation}
\overline{ l^{pref} } \equiv \int dl l P_{N}^{pref}(l) 
\label{lpref}
\end{equation}
represents a correlation length that remains finite in the high temperature phase as $N \to \infty$.
Since the second moment is also expected to be finite in the high temperature phase,
it is convenient to define the ratio
\begin{equation}
B_N(T)= \frac{\overline{ (l^{pref})^2 } }{ N \ \ \overline{ l^{pref} } }
\label{binder}
\end{equation}
which converge to $0$ in the high temperature phase, and
to a non-zero value at criticality and in the low temperature phase.
The results are shown on Fig. \ref{figbinder} (a) :
the critical temperature $T_c$ coincide with $T_1$
\begin{equation}
T_c = T_1 \sim 0.095
\end{equation}

The finite-size scaling of these data according to
\begin{equation}
B_N(T) \simeq {\cal B} \left(  (T-T_c) N^{1/{\nu}}  \right)
\label{binderfss}
\end{equation}
is consistent with the value
\begin{equation}
\nu \sim 2
\label{binderfssnu}
\end{equation}
as shown on Fig. \ref{figbinder} b

\begin{figure}[htbp]
\includegraphics[height=6cm]{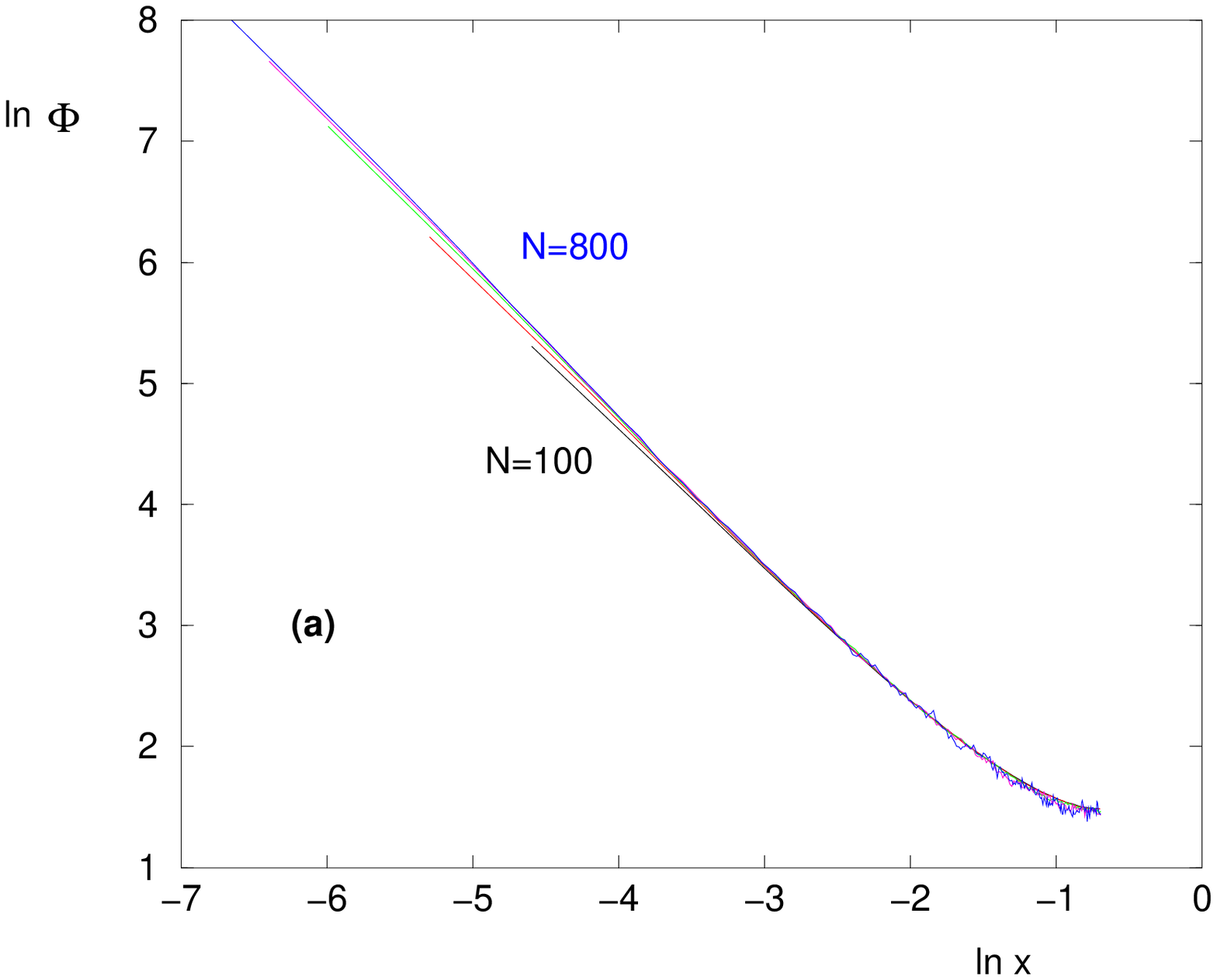}
\hspace{1cm}
\includegraphics[height=6cm]{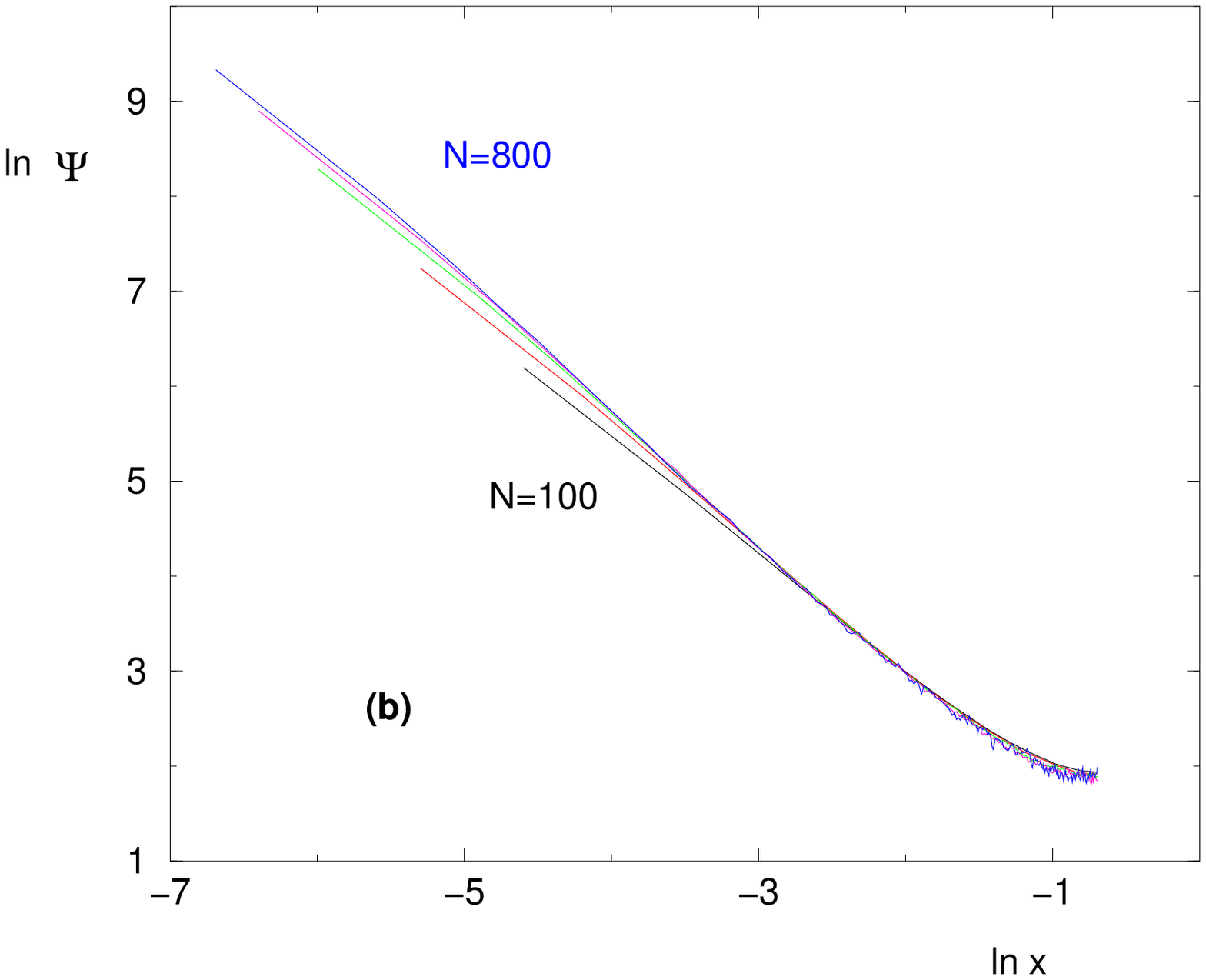}
\caption{(Color online) Scaling form of the probability distribution
$P_{N}^{pref}(l)$ :
log-log plot of $N^{\eta} P_{N}^{pref}(l)$ in terms of $x=l/N$ 
(a) for $T=0.05$ (low-T phase) the rescaling is done with $\eta=1.33$ (see Eq. \ref{prefllow})
(a) for $T_c \sim 0.095$ the rescaling is done with  $\eta_c=1.5$ (see Eq. \ref{prefltc}) }
\label{figdistrilpref}
\end{figure}

\begin{figure}[htbp]
\includegraphics[height=6cm]{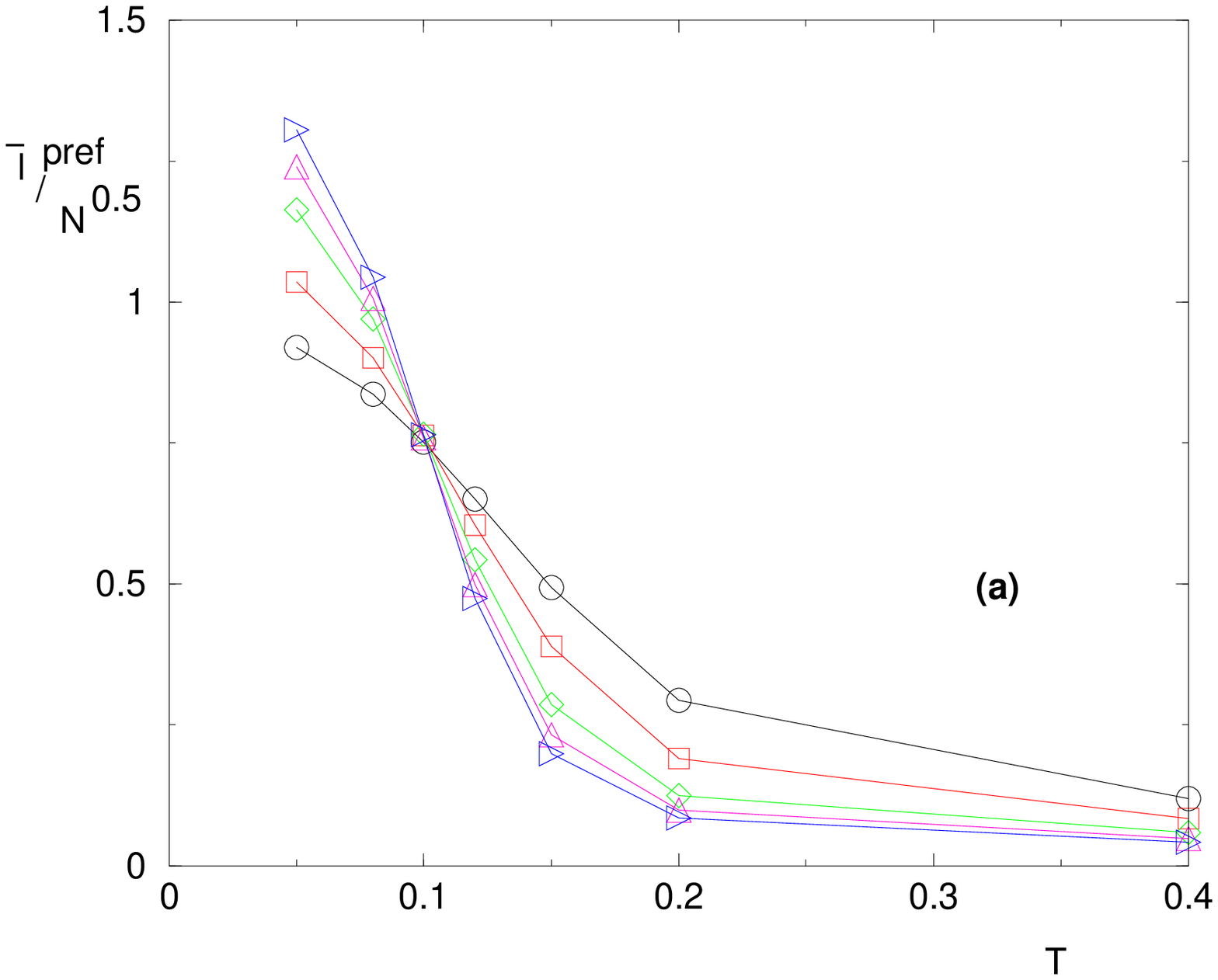}
\hspace{1cm}
\includegraphics[height=6cm]{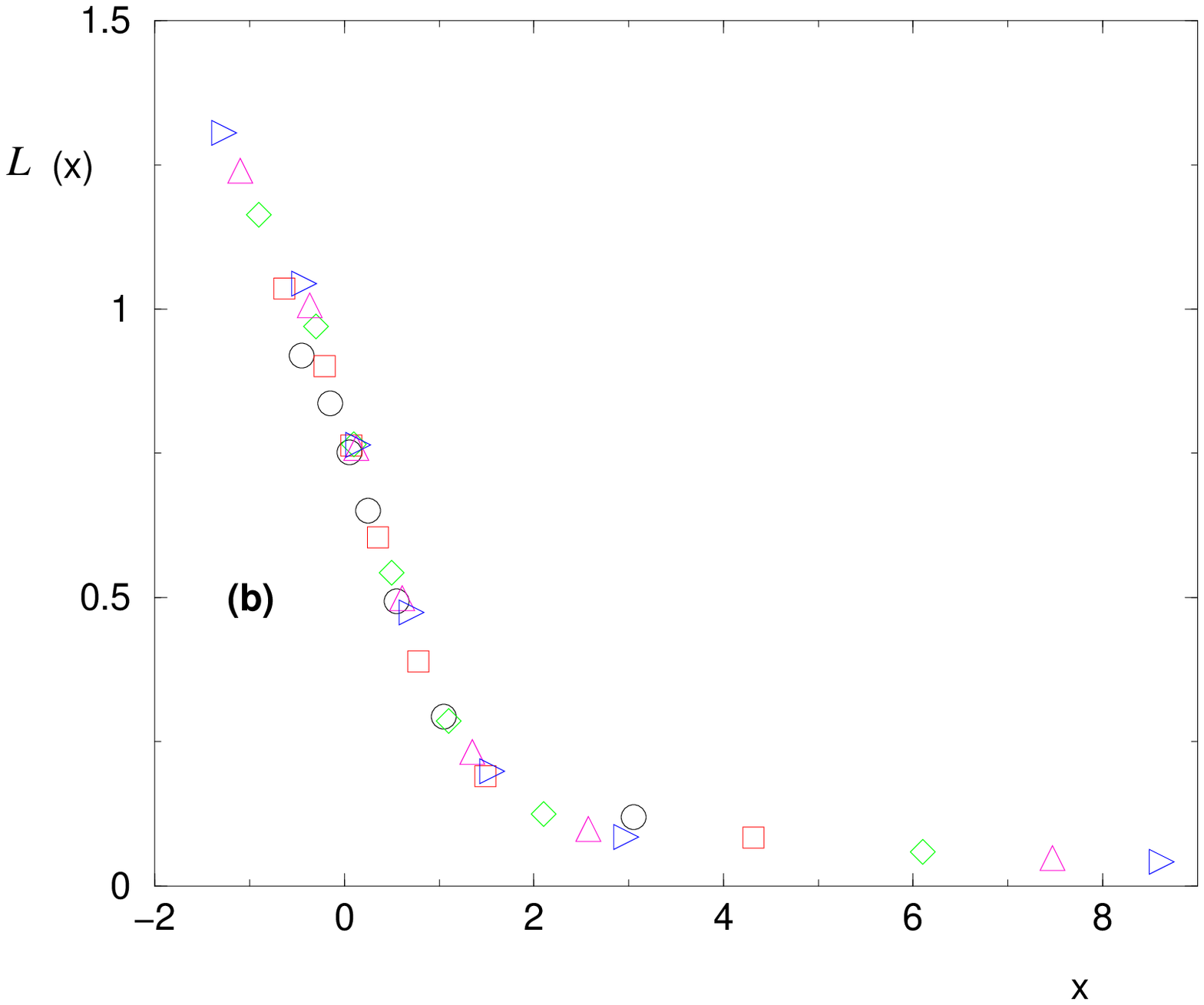}
\caption{(Color online) 
(a) $\overline{ l^{pref}(N) }/N^{0.5}$ as a function of $T$ for the sizes
 $N=100$
$(\bigcirc)$, $N=200$ $(\square)$, $N=400$ $(\lozenge)$, $N=600$
$(\vartriangle)$, $N=800$ $(\rhd)$
(b) finite size scaling of the same data in terms of the variable
$x=(T-T_c) N^{1/{\nu}}$
 (see  Eq. \ref{lpreffss}) with $T_c =0.095$ and $\nu=2$
 }
\label{figlprefcriti}
\end{figure}

We now discuss the behavior of $\overline{ l^{pref}} $ as a
function of $N$ for various temperatures.
For $T<T_c$, it grows as
\begin{equation}
\overline{ l^{pref} }  \sim N^{0.67}
\label{lpreflow}
\end{equation}
and the probability distribution $P_{N}^{pref}(l)$
follows the same scaling form as in the $T=0$ limit (Eq. \ref{preflt0})
\begin{equation}
\left[ P_{N}^{pref}(l) \right]_{T<T_c}   \sim  \frac{1}{N^{1.33}}
 \Phi \left( \frac{l}{N} \right) 
\label{prefllow}
\end{equation}
as shown on Fig. \ref{figdistrilpref} a.
At $T_c=T_1 \sim 0.095$, the first moment grows as
\begin{equation}
\overline{ l^{pref} }  \sim N^{0.5}
\label{lprefcriti}
\end{equation}
and the probability distribution $P_{N}^{pref}(l)$
follows the scaling form 
\begin{equation}
\left[ P_{N}^{pref}(l) \right]_{T_c}   \sim  \frac{1}{N^{1.5}}
 \Psi \left( \frac{l}{N} \right) 
\label{prefltc}
\end{equation}
as shown on Fig. \ref{figdistrilpref} b.
On Fig. \ref{figlprefcriti} a,
the rescaled variable $\overline{ l^{pref} }/N^{0.5}$ is shown as a
function of $T$ for various sizes : there is a crossing at $T_c$, and
the data follow the finite-size scaling behavior
\begin{equation}
\frac{\overline{l^{pref}} }{N^{0.5}} \simeq {\cal L} \left(  (T-T_c) N^{1/{\nu}}  \right)
\label{lpreffss}
\end{equation}
with $\nu \sim 2$ in agreement with the previous estimate of Eq. \ref{binderfssnu}.

\subsection{ Pair distribution $ \overline{P_{i,i+l}} $ and
height scaling }

\begin{figure}[htbp]
\includegraphics[height=6cm]{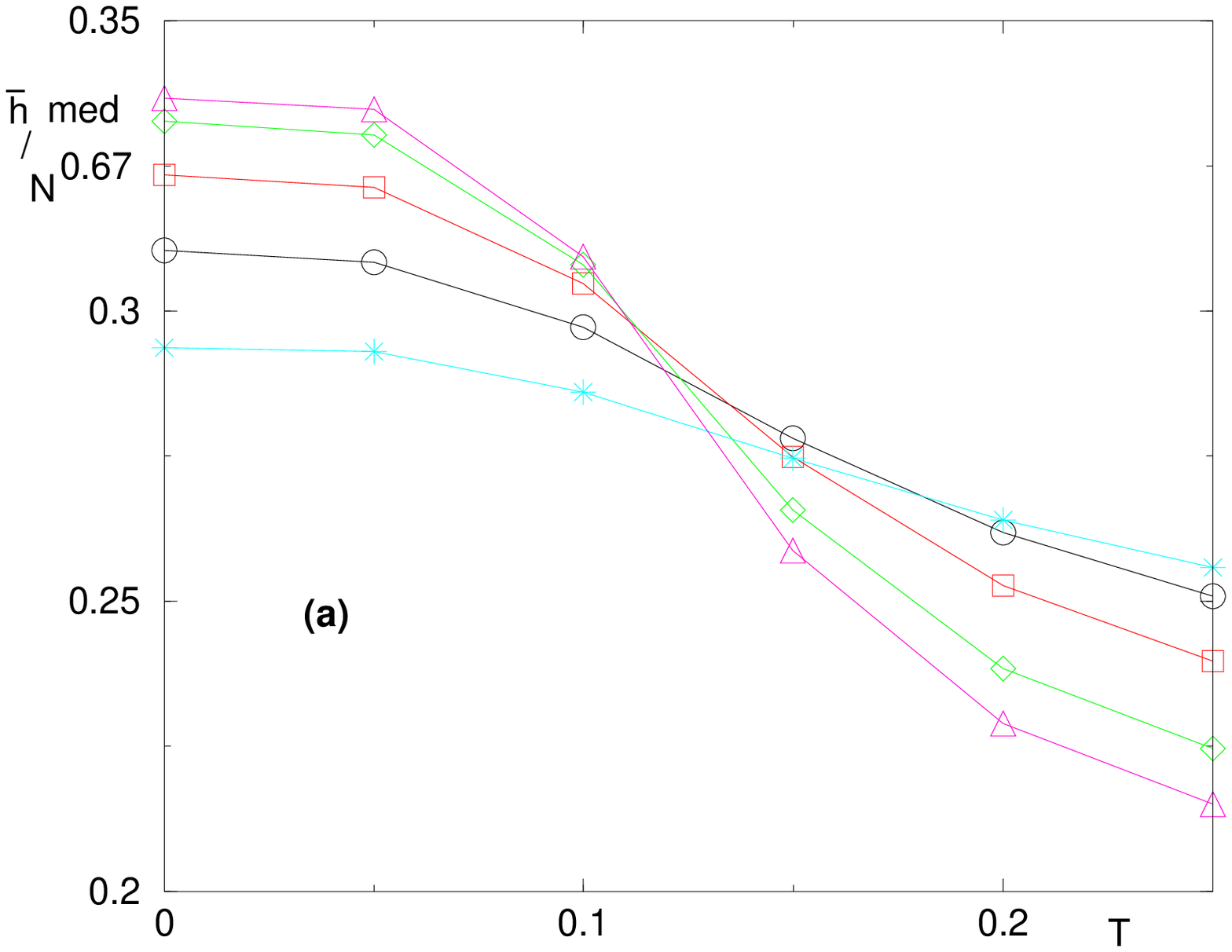}
\hspace{1cm}
\includegraphics[height=6cm]{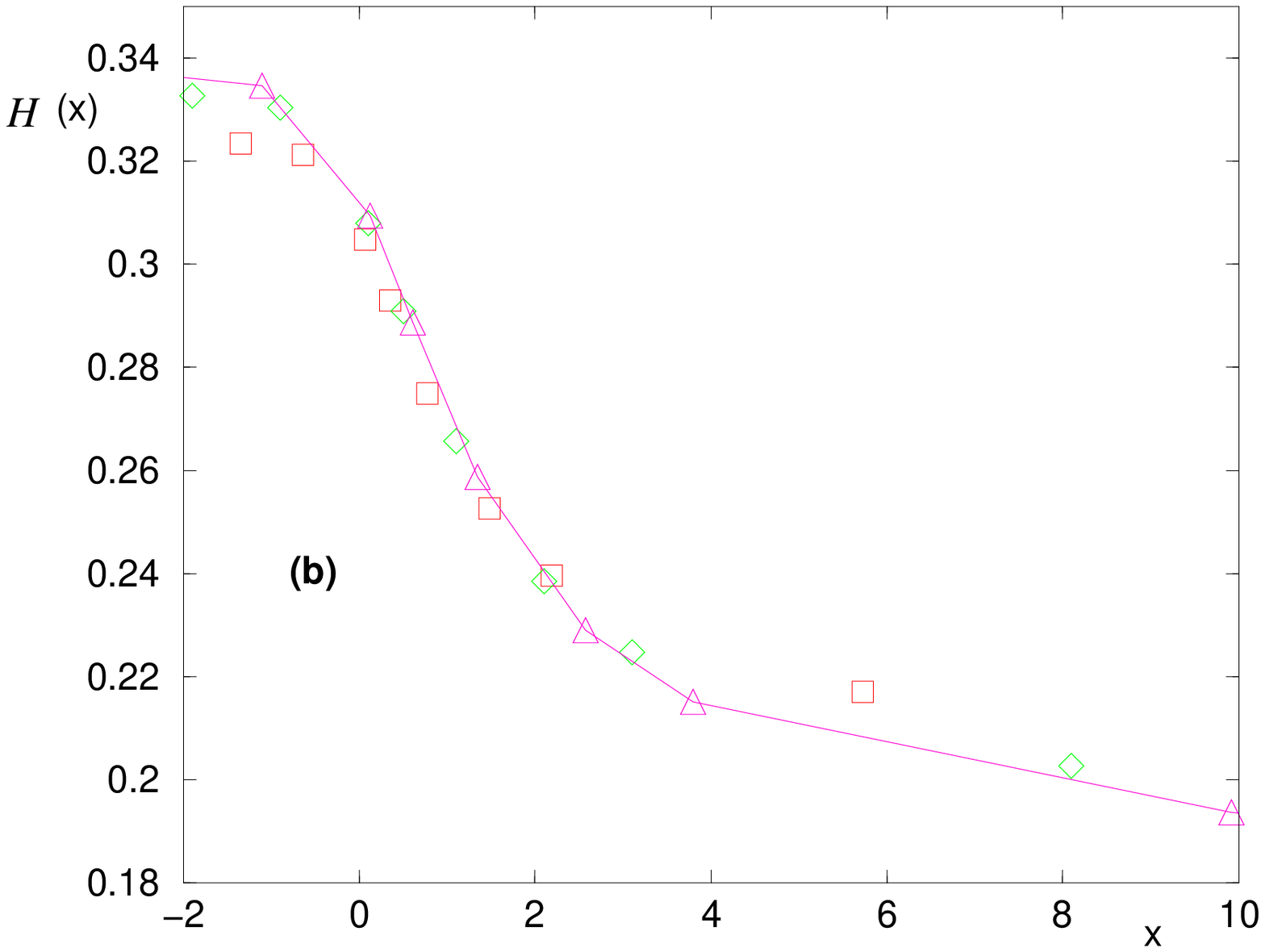}
\caption{(Color online) Height scaling : 
(a) the curves $ \overline{h_{med} }/N^{0.67}$
 of various sizes $N=50$ $(\ast)$,  $N=100$
$(\bigcirc)$, $N=200$ $(\square)$, $N=400$ $(\lozenge)$, $N=600$
$(\vartriangle)$ present crossings shifting towards $T_c$
(b) finite size scaling of the same data in terms of the variable
$x=(T-T_c) N^{1/{\nu}}$
 (see  Eq. \ref{hfss}) with $T_c =0.095$ and $\nu=2$ }
\label{fighauteur}
\end{figure}

We have measured the median height for each sample
\begin{equation}
h_{med} = \frac{1}{N} \sum_k <h_k>
\end{equation}
Its average over samples is directly related to the first moment
of the pair distribution $ \overline{P_{i,i+l}} $ ( see Eq. \ref{hk})
\begin{equation}
\overline{h_{med} }=   \sum_{l} l \  \overline{P_{i,i+l}} 
\label{hmed}
\end{equation}
We find that in the whole low-temperature phase and at $T_c$,
the roughness exponent is the same as at $T=0$ (Eq. \ref{ht0})
\begin{equation}
\overline{h_{med} } \sim N^{0.67} \ \ {\rm for } \ \ 0 \leq T \leq T_c 
\end{equation}
This is in agreement with Eq. \ref{zetaWie} quoted from \cite{Wie}.
Above $T_c$, the crossover towards the high-temperature roughness exponent
$\zeta=1/2$ (Eq. \ref{htinfty})
 is well described by the following finite-size scaling form
(see Fig. \ref{fighauteur} )
\begin{equation}
\frac{ \overline{h_{med} } }{N^{0.67}} \simeq
 {\cal H} \left(  (T-T_c) N^{1/{\nu}}  \right) \ \ 
{\rm for } \ \ T \geq T_c 
\label{hfss}
\end{equation}
with $\nu \sim 2$ in agreement with the previous estimates of
 Eqs \ref{binderfssnu} and \ref{lpreffss}.

Accordingly, we find that the pair distribution $ \overline{P_{i,i+l}}$
follows the $T=0$ finite-size scaling of Eq. \ref{pijt0}
in the whole low-temperature phase and also at $T_c$
\begin{equation}
\overline{ P_{i,i+l} (T_c) }  \sim  \frac{1}{N^{1.33}}
 \Phi \left( \frac{l}{N} \right) 
\label{pijtc}
\end{equation}

\subsection{ Overlap $ \overline{P^2_{i,i+l}} $ of large pairs $l$  }

\begin{figure}[htbp]
\includegraphics[height=6cm]{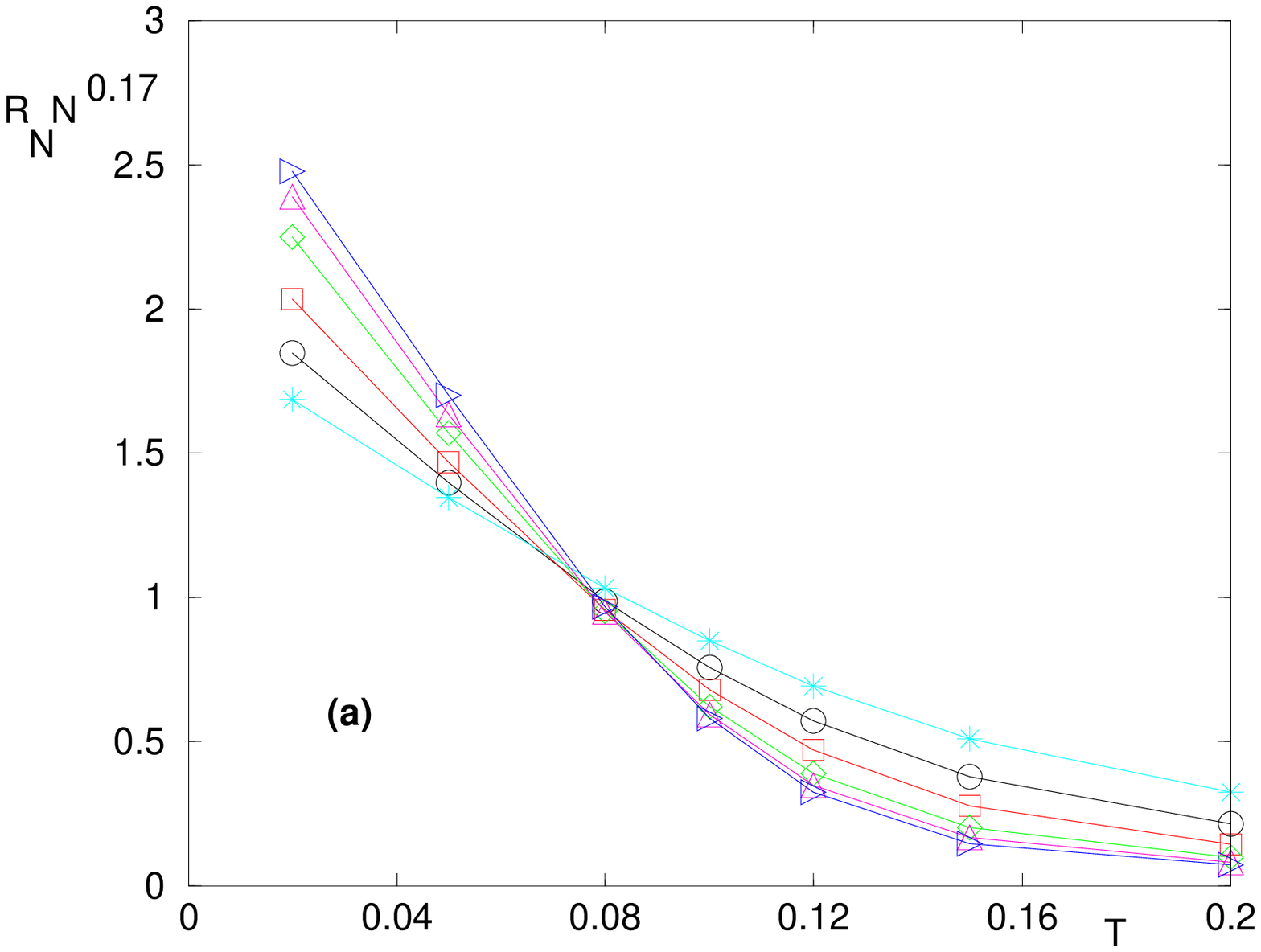}
\hspace{1cm}
\includegraphics[height=6cm]{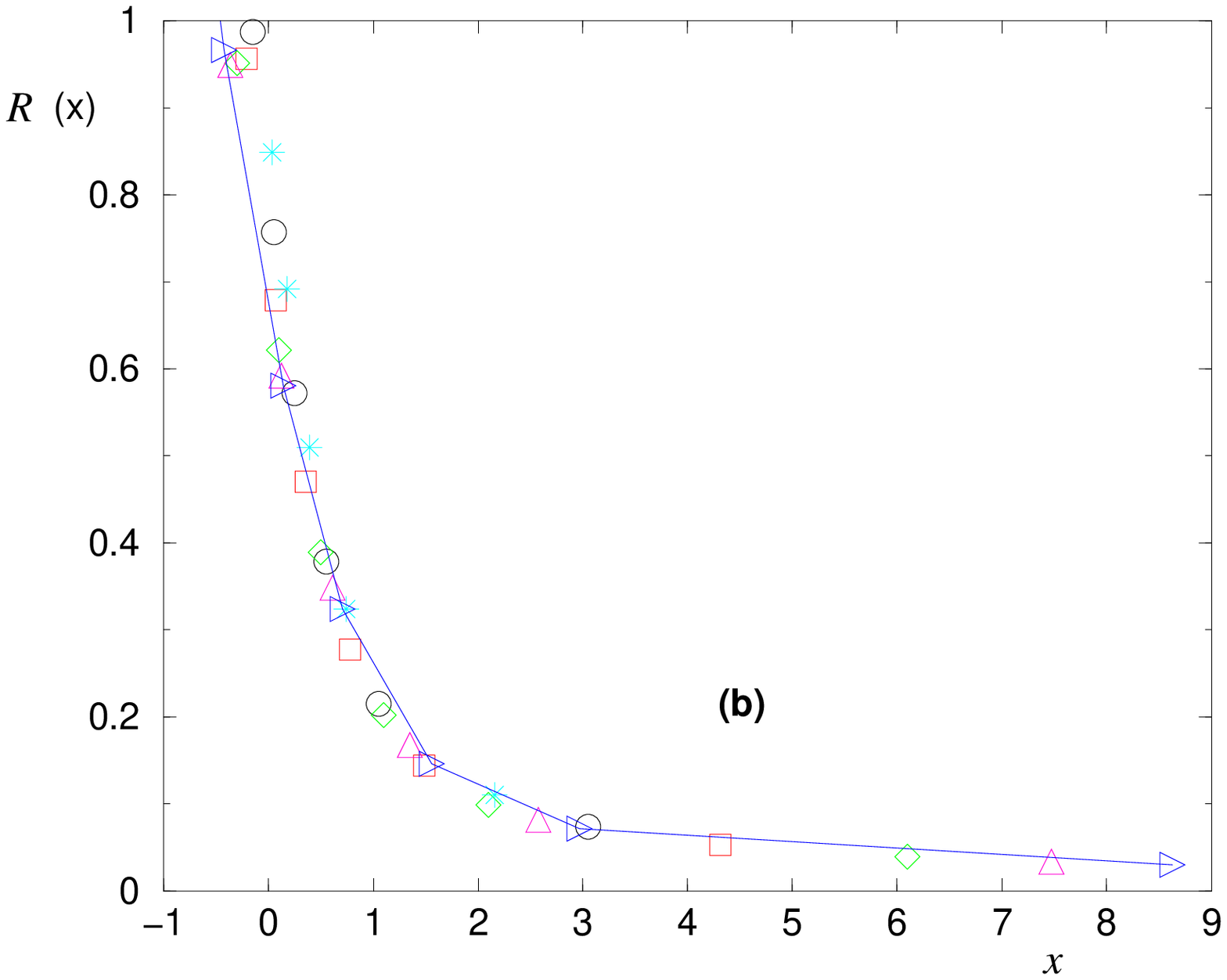}
\caption{(Color online) Critical behavior of the ratio $R_N(T)$
defined in Eq. \ref{ratiop2p1} :
(a) curves $N^{0.17}R_N(T) $ for the sizes  $N=50$ $(\ast)$ $N=100$
$(\bigcirc)$, $N=200$ $(\square)$, $N=400$ $(\lozenge)$, $N=600$
$(\vartriangle)$, $N=800$ $(\rhd)$
(b) finite size scaling of the same data in terms of the variable
$x=(T-T_c) N^{1/{\nu}}$
 (see  Eq. \ref{ratiop2p1fss}) with $T_c =0.095$ and $\nu=2$ }
\label{figratiop2p1}
\end{figure}

We have also measured the overlap $\overline{P^2_{i,i+l}}$
of large pairs. We find that the first moment scales as
\begin{eqnarray}
\int dl \  l \  \overline{P^2_{i,i+l} } && \sim N^{0.67} \ \ {\rm for} \ \ 0 \leq T < T_c  \\
\int dl \  l \  \overline{P^2_{i,i+l}} && \sim N^{0.5} \ \ {\rm for} \ T=T_c \\
\int dl \  l \  \overline{P^2_{i,i+l}} && \sim cte  \ \ {\rm for} \ T > T_c
\label{momentp2}
\end{eqnarray}
The scaling exactly at $T_c$ is distinct from the low-temperature phase
in disagreement with Eq. \ref{p2Wie} quoted from \cite{Wie},
but coincides with the scaling found above for $\overline{l^{pref}}$
(Eq. \ref{lprefcriti}). 
It is thus convenient to define the ratio
\begin{equation}
R_N(T)= \frac{ \int dl \  l \  \overline{P^2_{i,i+l}} }
{ \int dl \  l \  \overline{P_{i,i+l}} }
\label{ratiop2p1}
\end{equation}
which converge to $0$ in the high temperature phase, and
to a non-zero value in the low temperature phase.
Exactly at $T_c$, it is expected to decay
as $N^{0.5}/N^{0.67} = N^{- 0.17 }$. 
On Fig. \ref{figratiop2p1} (a), the curves $N^{0.17} R_N(T)$
present crossings that shift regularly towards $T_c$.
The finite-size scaling of these data according to
\begin{equation}
N^{0.17} R_N(T) \simeq {\cal R} \left(  (T-T_c) N^{1/{\nu}}  \right)
\label{ratiop2p1fss}
\end{equation}
with $T_c=0.095$ and $\nu=2$ is shown on Fig. \ref{figratiop2p1} b

\section{ Summary and conclusions}
\label{conclusion}

In this paper, we have analyzed the
 freezing transition of random RNA secondary structures
via the statistics of the pairing weights seen by a given monomer.
In analogy with L\'evy sums and Derrida's Random Energy Model
\cite{Der,Der_Fly}, we
have numerically computed the probability distributions 
$P_1(w)$ of the maximal weight, 
the probability distribution $\Pi(Y_2)$ of the parameter
$Y_2(i)= \sum_j w_i^2(j)$, as well as the average values of the moments
$Y_k(i)= \sum_j w_i^k(j)$. 
We have found two important temperatures $T_c<T_{gap}$.
For $T>T_{gap}$, the distribution $P_1(w)$ and $\Pi(Y_2)$ have a gap
and, accordingly, the moments $\overline{Y_k(i)}$ decay exponentially 
in $k$.
For $T<T_{gap}$, these moments decay with a power-law 
 $ \overline{Y_k(i)} \sim 1/k^{\mu(T)}$, and 
the distributions $P_1(w)$ and $\Pi(Y_2)$
present the characteristic Derrida-Flyvbjerg 
singularities at $w=1/n$ and $Y_2=1/n$
for $n=1,2..$. 
The most important singularities occur at $w=1$
with $P_1(w) \sim (1-w)^{\mu(T)-1}$ and $\Pi(Y_2) \sim (1-Y_2)^{\mu(T)-1}$.
 The exponent $\mu(T)$ increases with the temperature 
from the value $\mu(T=0)=0$ up to $\mu(T_{gap}) \sim 2$.
The study of spatial properties indicates that the critical temperature $T_c$
where the large-scale roughness exponent changes from the low temperature
value $\zeta \sim 0.67$ to the high temperature value $\zeta \sim 0.5$
corresponds to the exponent $\mu(T_c)=1$. 
The final picture is thus as follows : \\
(i) for $T<T_c$,  there exist frozen pairs of all sizes \\
(ii) for $T_c<T<T_{gap}$, there exist frozen pairs, but they are of 
finite-size. \\
(iii) for $T>T_{gap}$, even short pairs are not frozen anymore. \\
Finally, the finite-size scaling of various data are consistent
with a correlation length exponent $\nu \simeq 2$
that saturates the general bound $\nu \geq 2/d=2$
 of \cite{chayes} for phase transitions
in disordered systems.

In conclusion, the numerical study of the weight statistics 
appears as an interesting tool to clarify the nature
of low temperature phases existing in disordered systems.
In particular, we have shown that the frozen phase is characterized by
a temperature-dependent exponent $\mu(T)$ that governs the broadening
of the delta peak existing at $w=1$ at $T=0$.
We intend to study in a similar way other disordered models \cite{future}.

\appendix

\section{ Reminder on  L\'evy sums, the Random Energy Model
and Derrida-Flyvbjerg singularities}

\label{reminder}
\subsection{L\'evy sums when the first moment is infinite}

In this section, we recall some results on the weight statistics  \cite{Der}
for the case of L\'evy sums 
\begin{equation}
S_N = \sum_{i=1}^N x_i 
\end{equation}
of $N$ positive independent variables $(x_1,..x_N)$ distributed with
a probability distribution that decays algebraically
\begin{equation}
\rho(x) \opsimeq_{x \to +\infty} \frac{A}{x^{1+\mu} }
\label{levymu}
\end{equation}
 with $0<\mu<1$, i.e. when the first moment diverges
$<x>=+\infty$.  The sum $S_N$ then grows as $N^{1/\mu}$,
and the rescaled variable is distributed with a stable L\'evy
distribution \cite{levy}.
Another important property is that the maximal variable $x_{max}(N)$
among the $N$ variables $(x_1,...x_N)$ is also of order $N^{1/\mu}$,
i.e. the sum $S_N$ is actually dominated by the few biggest terms \cite{Der,levy} .
To quantify this effect, it is convenient to introduce the weights
\begin{equation}
w_i = \frac{x_i}{S_N}
\end{equation}
and their moments
\begin{equation}
Y_k=\sum_{i=1}^N w_i^k
\end{equation}
In particular, their averaged values in the limit $N \to \infty$
are finite for $0<\mu<1$ and reads \cite{Der} 
\begin{equation}
\overline{Y_k}^{Levy}= \frac{\Gamma(k-\mu)}{\Gamma(k) \Gamma(1-\mu)}
\label{yklevy}
\end{equation}
The density $f(w)$ giving rise to these moments
\begin{equation}
\overline{Y_k}^{Levy} = \int_0^1 dw w^k f(w)
\end{equation}
reads
\begin{equation}
f_{Levy}(w)= \frac{w^{-1-\mu} (1-w)^{\mu-1} }{ \Gamma(\mu) \Gamma(1-\mu)}
\label{densitew}
\end{equation}
and represents the averaged number of terms of weight $w$.
This density is non-integrable as $w \to 0$, because in the limit
 $N \to \infty$, the number of terms of vanishing weights diverges.
The normalization corresponds to 
\begin{equation}
\overline{Y_{k=1}}^{Levy} = \int_0^1 dw w f(w) =1
\end{equation}

More generally, correlations functions between $Y_k$ can also
be computed \cite{Der}, and the joint density of $K$ weights
reads \cite{Der_Fly}
\begin{equation}
f(w_1,...,w_K ) = \frac{\mu^{K-1} \Gamma(K)}
{\Gamma^K(1-\mu) \Gamma(K \mu)} \left( \prod_{i=1}^K w_i^{-1-\mu} \right)
 (1- \sum_{i=1}^K w_i)^{K \mu-1}
 \label{fKpoids}
\end{equation}

\subsection{Reminder on the Random Energy Model }

The Random Energy Model (REM) introduced in the context of spin glasses
\cite{rem} is defined by the partition function of $N$ spins
\begin{equation}
Z_N = \sum_{\alpha=1}^{2^N} e^{- \beta E_{\alpha}}
\end{equation}
where the $(2^N)$ energy levels are independent random variables
drawn from the Gaussian distribution
\begin{equation}
P_N(E) = \frac{1}{\sqrt{\pi N}} e^{ -\frac{E^2}{N}}
\label{PErem}
\end{equation}
It turns out that the low temperature phase $0<T<T_g$
of the REM \cite{Der,Der_Tou}
is in direct correspondence with L\'evy sums of index $0<\mu=\frac{T}{T_g}<1$ :
the weights
\begin{equation}
w_{\alpha} = \frac{ e^{- \beta E_{\alpha}} }{Z_N}
\end{equation}
have exactly the same moments $Y_k$ (Eq. \ref{yklevy}) and
the same density $f(w)$ (Eq. \ref{densitew}).
The explanation is that the lowest energy in the REM
are distributed exponentially 
\begin{eqnarray}
P_{extremal}(E) \opsimeq_{E \to -\infty} e^{ \gamma E}
\label{tailexp}
\end{eqnarray}
This exponential form that corresponds to the tail of the Gumbel distribution
for extreme-value statistics \cite{Gum_Gal,Bou_Mez},
immediately yields that the Boltzmann weight $x=e^{-\beta E}$ 
has a distribution that decays algebraically (Eq. \ref{levymu})
with exponent
\begin{eqnarray}
\mu= T \gamma
\end{eqnarray}
In the REM, the coefficient $\gamma$ in the exponential (Eq. \ref{tailexp})
is $\gamma=1/T_g$.

The link with the thermodynamics is that the entropy $S_N$
remains finite as $N$ grows and is given in terms of the 
weights by \cite{Gro_Mez}
\begin{equation}
S_N(T<T_g)= - \sum_{i=1}^N w_i \ln (w_i)
 = - \left [\partial_k \sum_{i=1}^N w_i^k \right]_{k \to 1}
= - \left [\partial_k Y_k \right]_{k \to 1}
= \Gamma'(1)- \frac{\Gamma' \left( 1- \mu(T) \right)}
{\Gamma \left( 1- \mu(T) \right)}
\end{equation}
and the corresponding specific heat $C_N(T<T_g)=T \partial_T S_N(T)$
then coincides with the finite-size result computed in \cite{rem}.
So the entropy per spin $S_N/N$ and the specific heat $C_N/N$
 vanish as $N \to \infty$ in the whole low-T phase. 
In the critical region $ T \to T_g^-$, 
the finite-size scaling behavior is
\begin{equation}
\frac{C_N(T)}{N} \propto \frac{1}{N (T_g-T)^2 }
\end{equation}

As a final remark, let us mention that in the mean-field
Sherrington-Kirkpatrick (SK) model
of spin-glasses, the same expressions of $Y_k$ (Eq. \ref{yklevy})
also appear \cite{Me_Pa_Vi,replica}, but with a different 
interpretation : the weights are those of the pure states.
As a consequence, the parameter $\mu(T)$, which is a complicated function
of the temperature, vanishes at the transition $\mu(T_c)=0$
(only one pure state in the high temperature state)
and grows as $T$ is lowered towards 
$\mu(T=0)$ of order $ 0.5$ \cite{SKzerotemp}.
This is in contrast with the REM model where $\mu(T)=T/T_g$
grows with the temperature from $\mu(T=0)=0$ (only one ground state)
to $\mu(T_g)=1$ at the transition, where the number 
of important microscopic states is not finite anymore.
Nevertheless, the expression (Eq. \ref{yklevy}) for the weights
of pure states means that the free-energy $f$ of pure states in the
SK model is distributed exponentially
\begin{eqnarray}
P(f) \opsimeq_{f \to -\infty} e^{ \gamma(T) f}
\end{eqnarray}
with a parameter $\gamma(T)= \mu(T)/T$.

\subsection{ Derrida-Flyvbjerg singularities}

In \cite{Der_Fly}, the statistics of the weights $w_i$
normalized to
\begin{eqnarray}
\sum_i w_i =1
\end{eqnarray}
for L\'evy sums with $0<\mu<1$ or equivalently of the REM or SK model
have been studied in details.
In particular, the probability distributions
$P_1(w)$ of the maximal weight
$w^{max}= max_i [w_i]$, $P_2(w)$
of the second maximal weight, and $\Pi(Y_2)$ of the parameter
$Y_2= \sum_i w_i^2$
present  
singularities at $w=1/n$ and $Y_2=1/n$
for $n=1,2..$ : this shows that 
all the weight is concentrated on a few terms.

The origin of these singularities is that 
the density of weights given in Eq. (\ref{densitew})
satisfy \cite{Der_Fly}
\begin{eqnarray}
f(w) && = P_1(w) \ \ {\rm for} \ \  \frac{1}{2} < w < 1
\nonumber  \\
f(w) && = P_1(w)+P_2(w) \ \ {\rm for} \ \  \frac{1}{3} < w < \frac{1}{2}
\nonumber  \\
f(w) && = P_1(w)+...+P_n(w)  \ \ {\rm for} \ \  \frac{1}{n+1} < w < \frac{1}{n}
\label{intervals}
\end{eqnarray}
For $w \to 1$, $P_1(w)$ thus presents the same singular behavior as
$f(w)$ of Eq. (\ref{densitew})
\begin{eqnarray}
 P_1(w=1-\epsilon ) \oppropto_{\epsilon \to 0} \epsilon^{\mu-1}
\end{eqnarray}
For $w \to 1/2$, the singularity of $P_1(w)$ 
comes from the two different expressions of Eqs \ref{intervals}
\begin{eqnarray}
P_1(w=\frac{1}{2}+\epsilon ) - P_1(w=\frac{1}{2}-\epsilon )
= f(w=\frac{1}{2}+\epsilon )- \left[ f(w=\frac{1}{2}-\epsilon )
-P_2(\frac{1}{2}-\epsilon) \right] \oppropto_{\epsilon \to 0}
 P_2(\frac{1}{2}-\epsilon)
\end{eqnarray}
For $1/3<w_2<w_1$, the joint probability that the two largest weights
are $w_1$ and $w_2$ is given by Eq. \ref{fKpoids} for $K=2$
\begin{equation}
f(w_1,w_2 ) = \frac{\mu }
{\Gamma^2(1-\mu) \Gamma(2 \mu)} w_1^{-1-\mu} w_2^{-1-\mu}
 (1- w_1-w_2)^{2 \mu-1}
\end{equation}
and thus $P_2(w_2)$ reads for $1/3<w_2<1/2$
\begin{equation}
P_2(w_2 ) = \int_{w_2}^{1-w_2}
 dw_1 f(w_1,w_2 )
 = \frac{\mu }
{\Gamma^2(1-\mu) \Gamma(2 \mu)} w_2^{-1-\mu} 
\int_{w_2}^{1-w_2} dw_1  w_1^{-1-\mu}
 (1- w_1-w_2)^{2 \mu-1}
\end{equation}
The singularity near $w_2 \to 1/2$ is thus of order
\begin{eqnarray}
P_2 \left(w_2= \frac{1}{2}-\epsilon \right) 
 \propto 
\int_{\frac{1}{2}-\epsilon}^{\frac{1}{2}+\epsilon} dw_1  
\left(\frac{1}{2}+\epsilon - w_1\right)^{2 \mu-1} \propto  \epsilon^{2 \mu}
\end{eqnarray}
So for $w_1,w_2 \to 1/2$, $P_1(w_1)$ and $P_2(w_2)$ have a singularity
of order $\epsilon^{2 \mu}$, i.e. there is no divergence,
in contrast with the singularity near $w \to 1$,
but there is an infinite slope for $0<\mu<1/2$.
More generally, the singularities of $P_1(w_1)$ at $w_1=1/n-\epsilon$
are weaker and weaker as $n$ grows according to \cite{Der_Fly}
\begin{eqnarray}
P_1 \left(w_1= \frac{1}{n}-\epsilon \right) 
 \propto  \epsilon^{2 \mu-2+n}
\end{eqnarray}
Similarly, for the distribution $\Pi(Y_2)$, the singularities are given by
\cite{Der_Fly}
\begin{eqnarray}
\Pi \left(Y_2= \frac{1}{n}-\epsilon \right) 
 \propto  \epsilon^{ n \left(\mu+\frac{1}{2}\right)-\frac{3}{2} }
\end{eqnarray}
In particular, near $Y_2 \to 1$, $\Pi(Y_2)$ exhibits the same
divergence as $P_1(w \to 1)$
\begin{eqnarray}
\Pi \left(Y_2= 1-\epsilon \right) 
 \propto  \epsilon^{ \mu-1}
\end{eqnarray}
and near $Y_2 \to 1/2$, the singularity
\begin{eqnarray}
\Pi \left(Y_2= \frac{1}{2}-\epsilon \right) 
 \propto  \epsilon^{ 2 \mu-\frac{1}{2}}
\end{eqnarray}
corresponds to an infinite slope for $0<\mu<\frac{3}{4}$.
We refer the reader to \cite{Der_Fly} for more details.
The shapes of $P_1(w)$, $P_2(w)$ and $\Pi(Y_2)$
can be found for $\mu=0.3$ and $\mu=0.1$ on Figs 3 and 4 of
\cite{Der_Fly} respectively.

Similar Derrida-Flyvbjerg singularities describe above
for the case of L\'evy sums or spin-glasses (REM or SK),
actually occur in many other contexts, such as
randomly broken objects \cite{Der_Fly,Kra},
in population genetics \cite{Higgsbio,Der_Jun,Dertrieste},
in random walk excursions or loops \cite{Der,Fra,Kan}.

\end{document}